\documentclass[a4paper,
11pt
]{article}

\usepackage[english]{babel}
\usepackage[a4paper,left=2cm,right=2cm,top=2.5cm,bottom=2.5cm]{geometry}

\usepackage[colorlinks=true, urlcolor=blue]{hyperref} 
\usepackage{bm}
\usepackage{authblk}

\usepackage{siunitx}
\usepackage[ruled,vlined]{algorithm2e}
\usepackage{algorithmic}
\usepackage{caption}
\usepackage{subcaption}
\usepackage{amssymb} 
\usepackage[dvipsnames]{xcolor}
\usepackage{amsmath, amssymb} 
\usepackage{booktabs}
\usepackage{rotating}
\usepackage{tikz}
\usepackage{colortbl}  
\usepackage{multirow}
\usepackage{array}
\usepackage{url}
\usepackage[square,numbers]{natbib}
\usepackage{enumitem}

\graphicspath{ {./images/} }

\title{
{
Projection-based model order reduction
for residence time distribution analysis of an
industrial-scale \\ 
continuous casting tundish}}

\author[1]{Harshith~Gowrachari\footnote{hgowrach@sissa.it}}
\author[3]{Mattia~Giuseppe~Barra\footnote{m.barra@danieli.com}}
\author[1]{Moaad~Khamlich\footnote{mkhamlic@sissa.it}}
\author[2]{Giovanni~Stabile\footnote{giovanni.stabile@santannapisa.it}}
\author[3]{Gianluca~Bazzaro\footnote{g.bazzaro@danieli.com}}
\author[1]{Gianluigi~Rozza\footnote{grozza@sissa.it}}

\affil[1]{\small Mathematics Area, mathLab, International School for Advanced Studies\\

Via Bonomea 265, 34136, Trieste, Italy.}

\affil[2]{\small Biorobotics Institute, Sant’Anna School of Advanced Studies\\

V.le R. Piaggio 34, 56025, Pontedera, Pisa, Italy.}

\affil[3]{\small Danieli Research Center, Danieli $\&$ C. S.p.A., Via Nazionale 41, 33042\\

Buttrio, Province of Udine, Italy.}

\date{} 

\begin{document}
\maketitle
    \begin{abstract}

The flow behavior in the continuous casting tundish plays a critical role in steel quality and is typically characterized via residence time distribution (RTD) curves. This study investigates the fluid flow behaviour in a single-strand tundish using numerical and experimental approaches. Full-order model (FOM) steady-state simulations were conducted under both isothermal and non-isothermal conditions to assess the influence of thermal buoyancy on the flow characteristics. The results show that buoyancy effects under non-isothermal conditions have a negligible impact on the overall velocity field. 

The converged flow fields serve as initial conditions for transient tracer transport simulations, enabling evaluation of RTD curves and volume partitioning. A Galerkin projection-based reduced-order model (ROM) is developed to efficiently derive RTD curves. Comparison of RTD curves from experiments, FOM simulations, and ROM predictions demonstrates strong agreement, with both computational approaches closely matching experimental data. 

A parameter-time dependent ROM is subsequently developed using Galerkin projection and operator interpolation. This enables efficient evaluation of RTD curves across varying parameter values with significantly reduced computational cost compared to full-order simulations. The ROM framework is well-suited for real-time analysis, design processes, optimization, and digital twin applications in metallurgical processes. \\

\textbf{Keywords:} continuous casting tundish; residence time distribution analysis; computational fluid dynamics; reduced-order model; Galerkin projection; digital twins.
\end{abstract}

    \section{Introduction}\label{sec:intro}

\begin{figure}[ht!]
    \centering
    \includegraphics[width=0.9\linewidth]{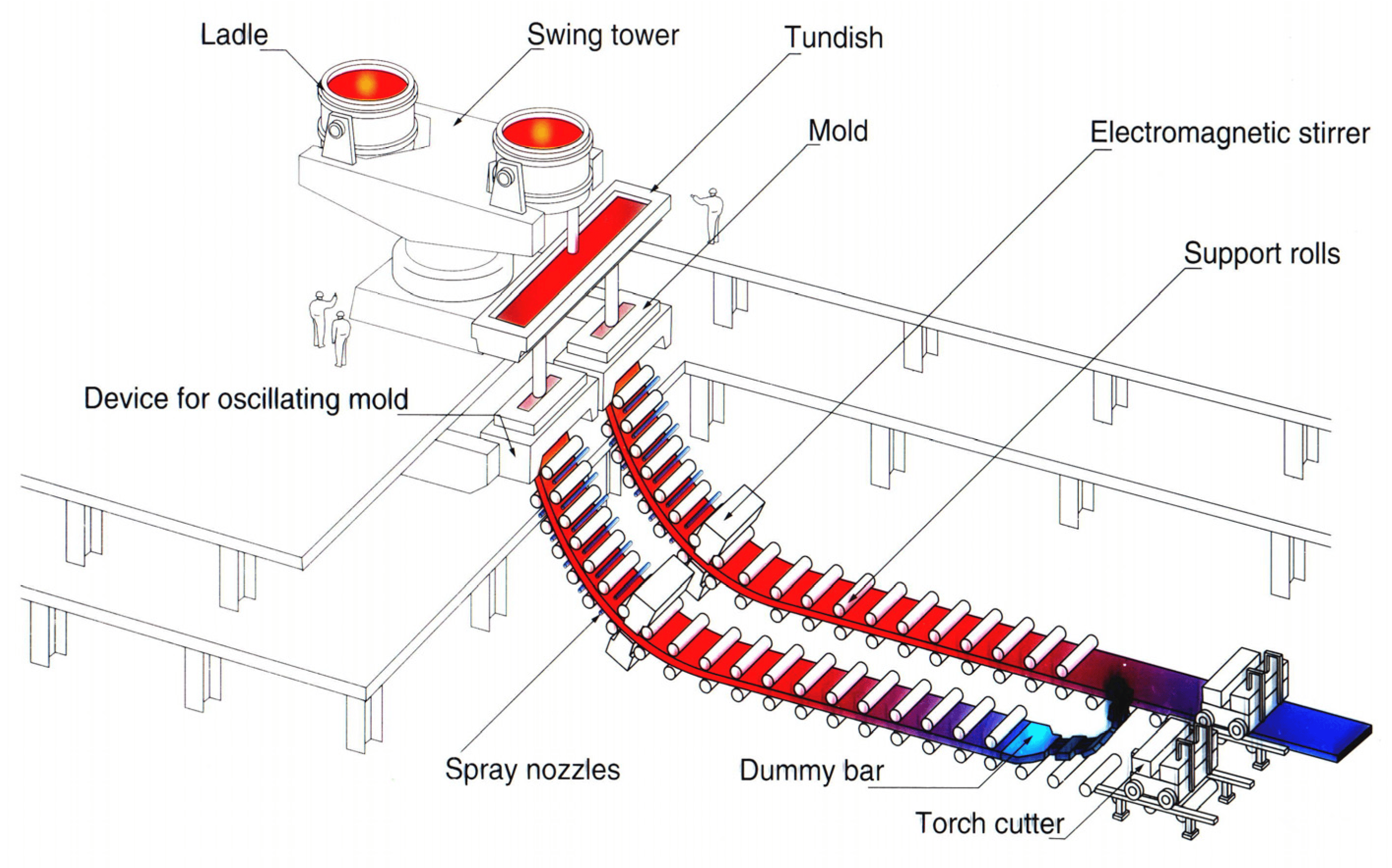}
    \caption{Schematic representation of a continuous casting plant showing the main components: ladle, tundish, molds in the primary cooling zone, and secondary cooling zone with roller system \cite{Guthrie2022}.}
    \label{fig:CCplant}
\end{figure}

The continuous casting process is the predominant method for solidifying molten steel into semi-finished products such as billets, blooms, and slabs, accounting for approximately 96$\%$ of worldwide steel production \cite{worldsteel2024}. Figure \ref{fig:CCplant} presents a schematic representation of a continuous casting machine. During the continuous casting of steel, the ladle containing molten steel coming from the production phase is placed on the rotating casting stand (ladle turret or swing tower) of the casting machine. Next, the steel flows through a submerged entry nozzle into oscillating molds in the primary cooling zone, where controlled solidification occurs \cite{Tkadlekov2020}. Below the molds, a system of rollers guides the partially solidified strand through the secondary cooling zone, where water jets from a series of nozzles complete the solidification process. \\

The tundish plays a critical role in the continuous casting process. It is a refractory-lined metallurgical vessel through which molten steel passes before entering the mold for solidification. Initially, the tundish was primarily viewed as a simple buffer reservoir between the ladle and the mold. It bridges the discontinuous process of secondary metallurgy (e.g., degassing, de-oxidation, alloying, temperature control, and cleanliness adjustment) with the continuous casting process, while distributing steel uniformly to multiple outlets \cite{Rueckert2009}. Over time, the paradigm has shifted, and the tundish is now regarded as a true metallurgical reactor, playing a very crucial role in ensuring the quality of the cast product. \\

The growing demand for high-quality steel has heightened awareness among steelmakers regarding product cleanliness requirements \cite{Zhang2003}. Steel cleanliness is primarily determined by the concentration of non-metallic inclusion particles that can be either endogenous or exogenous in origin. If not effectively removed during processing, these inclusions can significantly degrade the quality of the final steel product \cite{Tkadlekov2020}. The composition, size, shape, quantity, and distribution of non-metallic inclusions significantly influence steel performance, affecting properties such as strength, toughness, fatigue resistance, and corrosion resistance \cite{Ren2022}. Methods for minimizing non-metallic inclusions primarily involve controlling the composition of steel, slag, and refractory materials. Additionally, regulating fluid flow patterns within each vessel during refining and continuous casting is crucial \cite{Zhang2020}. \\

Since the tundish is the final stage in the continuous casting process, where the content of non-metallic inclusions in the steel melt can be substantially modified \cite{Zhang2020}, investigating and optimizing the fluid flow within this vessel is critical to ensuring product quality. \\ 

To quantitatively assess tundish performance, a widely used method for analyzing molten steel flow is the stimulus–response technique \cite{Wang2022}. In this approach, the stimulus or input involves introducing a tracer material into the steel stream entering the tundish. Although various tracer input types may be applied, step and pulse inputs are the most commonly employed. The response, or output, is the concentration–time profile measured at the tundish outlet \cite{Sahai2007}, commonly referred to as the residence time distribution (RTD) curve. \\

The stimulus–response technique is suitable for analyzing both single-strand (single outlet) and multi-strand (multiple outlets) tundishes. In the case of multi-strand tundishes, the overall RTD curve is determined by calculating the weighted average of the individual RTD curves obtained at each outlet. The weighting factors correspond to the flow rates through each outlet, which may differ due to variations in flow distribution. \\

The RTD approach provides valuable insights into the flow characteristics of molten steel within the tundish. Among the most important flow phenomena that RTD analysis can identify is the presence or absence of flow shortcuts, the preferential paths through which steel reaches the outlet with minimal residence time. Steel that passes through preferential paths retains a high concentration of non-metallic inclusions, as the low residence time prevents sufficient flotation and removal of impurities. The presence of such flow patterns adversely affects steel cleanliness by promoting the discharge of inclusions from the tundish. In RTD curves, these shortcuts are typically indicated by multiple early peaks in concentration or by a dominant peak occurring at very low residence times. \\

Another key parameter derived from RTD analysis is the dead volume fraction within the tundish. Dead volume refers to regions where the steel residence time exceeds twice the theoretical residence time, which is defined as the ratio of tundish volume to the volumetric flow rate at the inlet. These zones are characterized by low velocity, often exhibiting recirculating flow patterns. A high dead volume fraction can lead to several operational issues, including steel freezing (partial solidification) due to prolonged residence time, and reduced inclusion removal efficiency, as inclusions tend to become trapped within these stagnant regions. \\

\begin{figure}[ht!]
    \centering
    \includegraphics[width=0.75\linewidth]{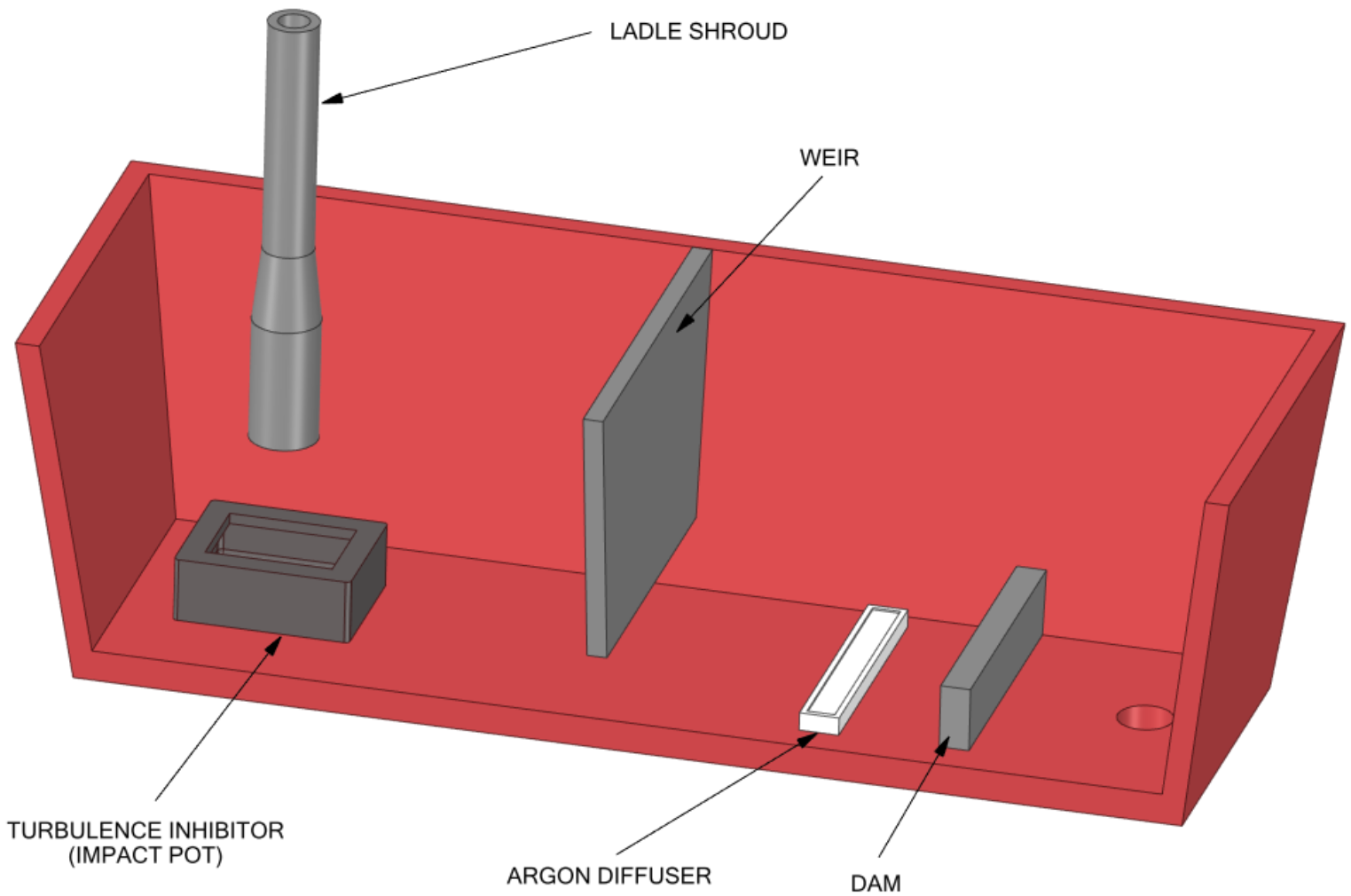}
    \caption{Schematic representation of a single-strand tundish showing various flow control devices including impact pot, weir, dam, and argon diffuser designed to optimize molten steel flow patterns and improve metallurgical performance.}
    \label{fig:tundish_schematics}
\end{figure}

A detailed analysis of the RTD curve provides valuable insights into the effectiveness of tundish design. If necessary, it also offers guidance on modifying the tundish design to improve the steel flow pattern. Various flow control devices, such as turbulence inhibitors (impact pots), weirs, dams, and argon diffusers, are commonly located within the tundish volume to enhance tundish performance. Figure \ref{fig:tundish_schematics} presents a simplified schematic of a single-strand tundish, illustrating various flow modifiers that could be placed inside the tundish volume. \\

RTD curves for tundishes can be obtained through both numerical and physical simulations. Physical experiments are typically carried out using full-scale or scaled-down water models, where molten steel flow is simulated with water, and a conductive NaCl solution is commonly used as a tracer \cite{Sahai2007, Dinda2024}. The physical properties of the tracer, along with the injection method, can significantly influence the experimental results \cite{Liu2022, Damle1995, Brard2020}. Factors such as tracer concentration, density, quantity, and type all affect the accuracy and reliability of the RTD measurements. While lower tracer concentrations and smaller quantities are generally preferred to minimize flow disturbance, the sensitivity of the concentration measurement devices imposes a practical lower limit on the tracer dosage \cite{Liu2022, Brard2020, Chen2012}. \\

Accurate simulation results, including the RTD curve, are contingent upon the proper selection of the turbulence model used to characterize the liquid metal flow. Model verification is typically conducted by comparing simulation results with data obtained from water model experiments \cite{Liu2022}. The test facility used for the water model experiment, located at the Danieli Research Center, is discussed in Section \ref{sec:EXPT}. \\

While high-fidelity simulations provide accurate RTD predictions, their computational expense limits their practical application in parametric studies, optimization, and real-time monitoring. To address this challenge, projection-based model order reduction methods offer a systematic framework for reducing the computational cost associated with full-order simulations. These methods aim to construct ROMs that capture the essential dynamics of the FOMs while operating in a low-dimensional subspace. Among the most widely used dimensionality reduction techniques in model order reduction (MOR), proper orthogonal decomposition (POD) has gained significant traction for its ability to extract the most energetic modes from high-fidelity full-order simulations. \\

In the POD-Galerkin method, the governing equations are projected onto the reduced subspace, which is spanned by POD basis functions, and is a widely adopted approach in reduced-order modelling. Hall et al. \cite{Hall2000} employed the snapshot {POD} method to model unsteady aerodynamic flows in both transonic and subsonic regimes, examining configurations such as an isolated airfoil and a cascade of flat plate airfoils. Their work demonstrated the ability of {POD}-based {ROM} to deliver accurate predictions, highlighting their potential for use in active flow control. In particular, POD-Galerkin {ROM} has found widespread application across various domains, including optimal control, design optimization, data assimilation, and real-time query systems. Ravindran \cite{Ravindran2000} developed a POD-Galerkin {ROM} for optimal fluid flow control in channel problems, highlighting both the model's accuracy in short-time prediction and its ability to deliver significant computational speed-ups. Such characteristics are essential for time-sensitive applications like feedback control. Similarly, Bourguet and Braza \cite{Bourguet2007} investigated compressible, unsteady transonic flows over a NACA0012 airfoil using a POD-Galerkin {ROM}. Their model captured key flow instabilities, such as von Kármán vortex shedding and buffeting, with excellent agreement to full-order simulations, although the evaluation of non-linear terms remained computationally demanding. \\

Further advancements in {ROM} development include notable efforts using finite volume FOMs of the Navier-Stokes equations (NSE), as demonstrated in the foundational studies \cite{Lorenzi2016, stabile2018finite, stabile2019reduced, Haasdonk2008}. Ballarin et al. \cite{Ballarin2016} proposed a monolithic POD-Galerkin approach for fluid-structure interaction problems. Also in \cite{Ballarin2014} presented a stable POD-Galerkin ROM tailored for the unsteady, incompressible, and parameter-dependent {NSE}. Another contribution by Ballarin et al. \cite{Ballarin2016haemodynamics}, the use of POD-Galerkin ROMs in the context of haemodynamics, further highlights the versatility of model reduction techniques for parameterized PDEs.\\

Several efforts have been made to extend POD-Galerkin ROMs to non-isothermal problems. An early attempt is reported in \cite{aling1997nonlinear}, where the authors developed a POD-Galerkin ROM for modelling the temperature field within a rapid thermal processing chamber, focusing on a two-dimensional steady-state case. In \cite{Alonso2009}, Alonso et al. combined {POD} with a genetic algorithm to construct a ROM for heat transfer in a backward-facing step flow. A one-dimensional conduction problem was addressed in \cite{Raghupathy2009} using a POD-Galerkin approach, and further studies on the heat conduction equation were carried out in \cite{Wang2012}and \cite{Han2014}. Natural circulation was investigated in \cite{Li2013} by developing an {ROM} from a {FOM} based on the coupled Navier-Stokes and energy equations. However, the resulting reduced model only allowed variations in the two-dimensional temperature field, keeping the flow field fixed, thus limiting its applicability to small perturbation scenarios in temperature control. In \cite{Pasetto2011}, a POD-Galerkin ROM was proposed for groundwater flow problems subject to spatially distributed stochastic inputs. The authors generated snapshots in the probability space, leading to a reduced-order Monte Carlo framework (ROMC). An alternative {ROM} approach is explored in \cite{Mller2011} for uncertainty propagation in porous media, where the Karhunen-Loève decomposition (equivalent to {POD}) and polynomial chaos expansion with Smolyak sparse grids were applied. The {POD} method was also employed in \cite{Li2011} for two-dimensional solute transport problems. \cite{Busto2020} presented a POD-Galerkin ROM for weakly coupled Navier-Stokes and heat transport equations, using a hybrid finite element–finite volume discretization. Finally, Georgaka et al. \cite{SokratiaGeorgaka2020} developed a POD-Galerkin reduced-order model for the three-dimensional unsteady Navier–Stokes equations minimally coupled with the heat transport equation, parametrizing both the boundary condition and physical parameter variations.\\

In metallurgical applications such as {RTD} analysis in tundish systems, projection-based model order reduction enables efficient simulation of unsteady transport phenomena. RTD analysis is essential for assessing flow behavior and mixing characteristics within the tundish, both of which directly affect steel cleanliness and product quality. High-fidelity simulations of {RTD} based on the Navier–Stokes and scalar transport equations offer accurate representations of these transport phenomena. However, these models are computationally expensive, especially for parametric studies or real-time applications. By projecting the governing equations onto a reduced basis subspace, ROMs yield fast and accurate approximations of scalar concentration fields. This approach facilitates detailed {RTD} characterization across varying flow configurations and operational conditions, making it particularly valuable for design optimization, process monitoring, and digital twin development in continuous casting. \\

In this study, we adopt the framework proposed by Khamlich et al. \cite{Khamlich2023} to construct a projection-based ROM tailored for RTD analysis of an industrial-scale continuous casting tundish. To compute the convective flow fields from steady-state simulations, projection-based model order reduction can be applied under both isothermal and non-isothermal conditions, following the procedures outlined in \cite{stabile2018finite, stabile2019reduced, SokratiaGeorgaka2020}. In the non-isothermal case, the treatment of non-linear terms poses additional computational challenges, often requiring stabilization techniques or hyper-reduction strategies. \\

In this work, we propose an intrusive projection-based ROM framework for RTD analysis in an industrial-scale continuous casting tundish, building on the methodology presented in \cite{Khamlich2023}. Specifically to enable rapid online evaluation of the reduced operators, we incorporate an operator approximation strategy based on regression models, whose accuracy is assessed in Section~\ref{sec:results}.

\vspace{0.5em}

The main contributions of this work are as follows:

\begin{itemize}

    \item We develop an intrusive projection-based ROM for RTD analysis in an industrial-scale, three-dimensional continuous casting tundish.

    \item We validate the proposed ROM against experimental data and high-fidelity FOM simulation results, demonstrating its accuracy.

    \item We show that the ROM provides reliable predictions for unseen parameters, with good agreement with FOM results. Additionally, we assess the effectiveness of a regression-based operator approximation strategy (Radial Basis Functions) for the rapid online evaluation of reduced operators in an industrial setting.

    \item The proposed physics-based ROM framework is efficient and scalable, enabling real-time process monitoring, optimization, and design-space exploration for continuous casting operations, thus offering a practical tool to enhance steelmaking processes.

\end{itemize}

This paper is organised as follows. In Section \ref{sec:EXPT}, we describe the experimental setup that serves as a basis for validation; in Section \ref{sec:problem}, we formulate the physical problem under consideration; in Section \ref{sec:FOM}, we detail the full-order model, including the governing equations, turbulence modelling, the Boussinesq approximation, and species transport; in Section \ref{sec:num_method}, we outline the numerical methods employed; in Section \ref{sec:ROM}, we present the projection-based reduced-order modelling framework, structured into offline and online stages; in Section \ref{sec:num_setup}, we describe the numerical setup, including the computational domain and boundary conditions; the results and discussion are presented in Section \ref{sec:results}; and finally, conclusions and perspectives are discussed in Section \ref{sec:conclusion}.

    \section{Experiment}\label{sec:EXPT}

\begin{figure}[ht!]
    \centering
    \includegraphics[width=1\linewidth]{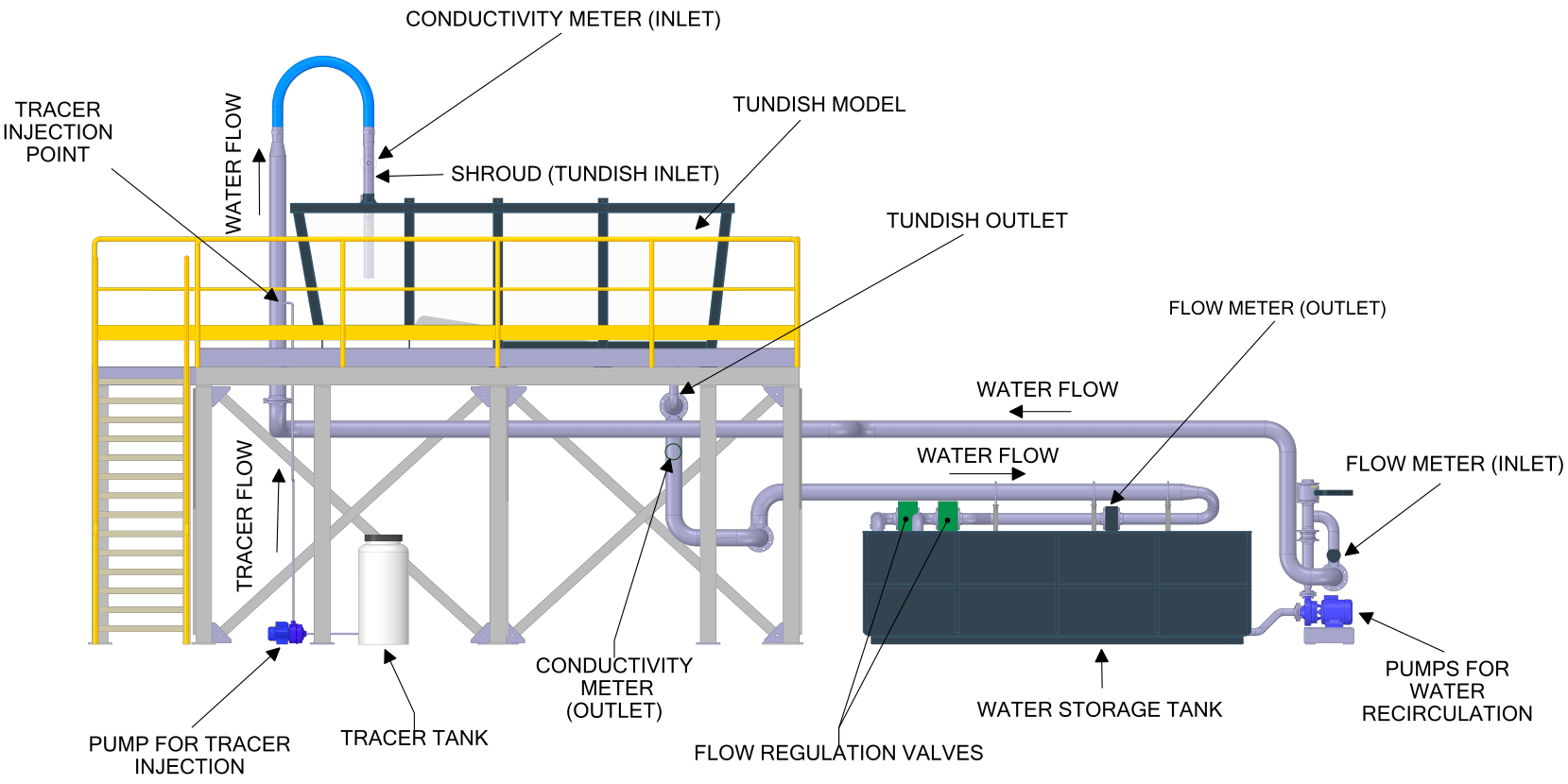}
    \caption{ Test facility dedicated to water model experiments on tundish models. The test facility depicted is located at the Danieli Research Center in Buttrio, Province of Udine, Italy. }
    \label{fig:expt_Setup}
\end{figure}

The experimental setup used to measure the RTD curve of the tundish is illustrated in Figure \ref{fig:expt_Setup}. The experimental setup consists of a large storage tank that supplies water to the system. Two variable-speed pumps draw water from the tank and deliver it to the tundish inlet at the desired flow rate. The outlet flow is gravity-driven and regulated by two control valves installed along the outlet pipeline, which returns the water to the storage tank. The test facility is equipped with a control system that enables precise regulation of the water flow rate and maintenance of a constant water level in the tank. It also facilitates the controlled injection of tracer material into the water flow.  \\

The tracer used in the experiments consists of a salt solution prepared by dissolving 10 kg of NaCl in 200 liters of industrial water. This solution is stored in a dedicated tank positioned beneath the tundish structure. A dedicated pump ensures the injection of the tracer into the system. The stimulus–response technique is employed in the experiment. The tracer is injected through the shroud, and its concentration at the inlet and outlet is continuously monitored by two conductivity meters.   \\

The experiments were conducted on a full-scale Plexiglass model replicating the actual tundish. A unitary scaling factor was applied to preserve both Froude and Reynolds number similarities. Although the Reynolds number does not match exactly, the discrepancy remains within 10$\%$ \cite{Sahai2007}, primarily due to the difference in kinematic viscosities between water at room temperature and molten steel. However, this deviation is considered negligible for practical purposes. As a result, the Reynolds similarity criterion is reasonably satisfied. The use of a full-scale model ensures both geometric and dynamic similarity with the actual tundish, allowing the volumetric flow rates in the water model to correspond directly to those in the real plant casting process with molten steel. \\

Once the operational parameters (water flow rate and water level within the tundish) are set and the pumps are activated, a waiting period is required before commencing the test. This allows the system to reach steady-state conditions and permits the dissipation of transient effects arising from the tundish filling process. Following this stabilization period, tracer injection is initiated for 30 seconds. During the test, the data acquisition system continuously monitors the tracer concentration at both the inlet and outlet. The test is considered complete when the conductivity meter readings at both locations stabilize and approximate a steady state, typically occurring 40 to 50 minutes after tracer injection. \\

\begin{figure}
    \centering
    \includegraphics[width=0.8\linewidth]{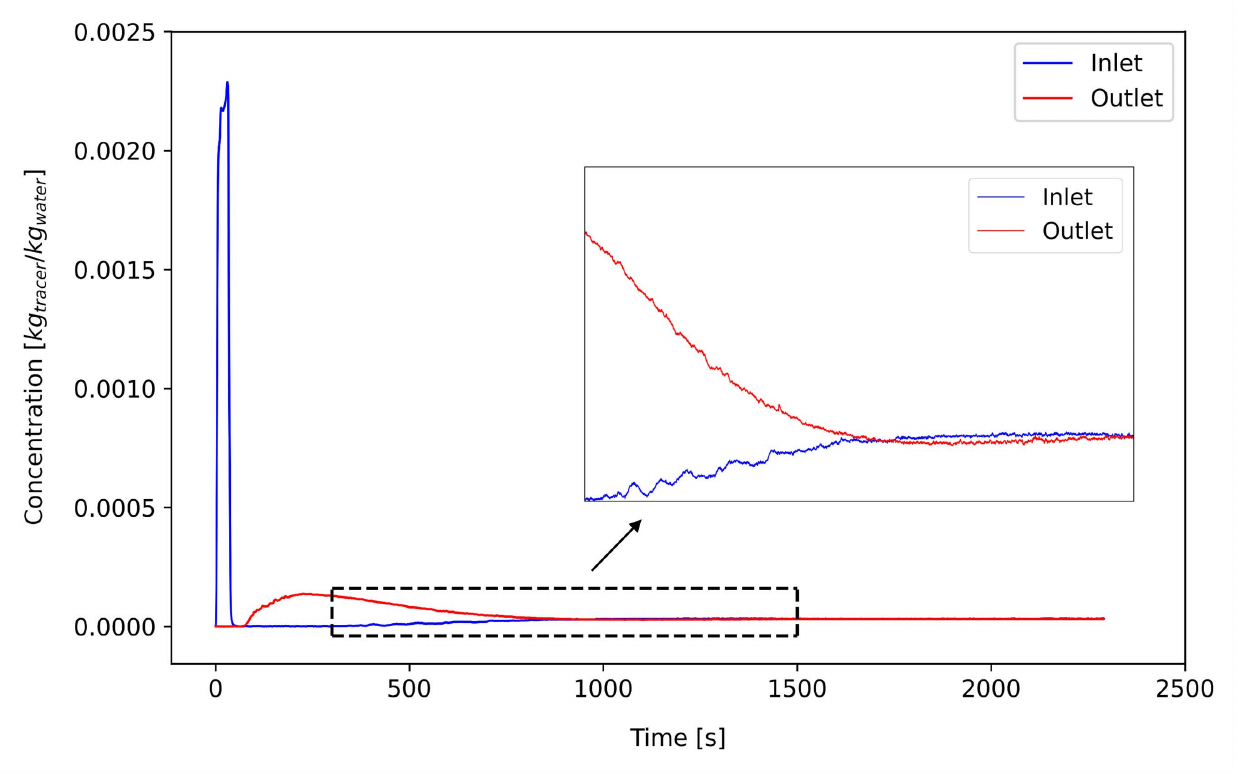}
    \caption{Concentration-time curves measured at the inlet and outlet of a single-strand tundish, highlighting the effect of the recirculation of the tracer inside the system.}
    \label{fig:Concentration-timecurves}
\end{figure}

To obtain the real RTD curve of the tundish, which represents the system's impulse response, the tracer must be circulated only once, as opposed to continuously, as is done in the closed-loop water circuit of the facility. Figure \ref{fig:Concentration-timecurves} shows an example of the concentration-time curves measured at both the inlet and outlet of a single-strand tundish tested in the water model facility, highlighting the effect of tracer recirculation, which leads to an increase in tracer concentration at the inlet. \\

Starting with the RTD curve measured at the water model facility, the true RTD curve of the tundish can be reconstructed through a deconvolution process. For a linear, time-invariant system (LTI), the system's response \( y(t) \) to a generic input \( x(t) \) is represented by the convolution integral of the impulse response \( g(t) \) and the input \( x(t) \):

\begin{equation}
    y(t) = \int_{0}^{t} g(\tau) x(t-\tau) \, d\tau .   
\end{equation}

The discretized form of the convolution integral is given by:

\begin{equation}
    y(t) \cong \sum_{0}^{n} g(\tau_i) x(t-\tau_i) \, \Delta\tau   .
    \label{eqn:discrete-conv}
\end{equation}

\begin{figure}[ht!]
    \centering
    \includegraphics[width=1\linewidth]{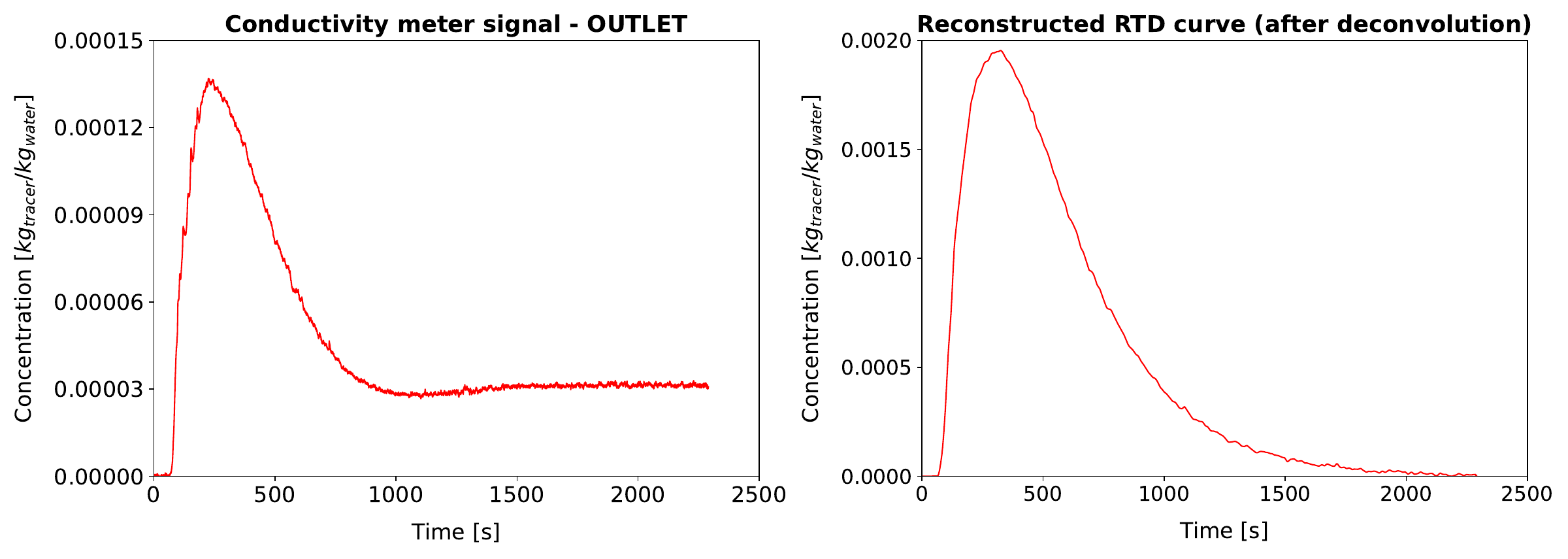}
    \caption{Deconvolution process to obtain the actual RTD curve of the tundish.}
    \label{fig:Deconvolution}
\end{figure}

In this context, \( y(t) \) represents the signal from the sensor measuring tracer concentration at the tundish outlet, while \( x(t) \) represents the signal from the sensor at the inlet. The function \( g(t) \) denotes the impulse response of the system, which corresponds to the true RTD curve of the tundish. By numerically solving (\ref{eqn:discrete-conv}) for \( g(t) \) (via the deconvolution process), the impulse response of the tundish can be reconstructed, thereby obtaining its RTD curve. Figure \ref{fig:Deconvolution} illustrates an example of the application of the deconvolution process to reconstruct the actual RTD curve of the tundish.

\section{Physical problem}\label{sec:problem}

\begin{figure}[ht!]
    \centering
    \includegraphics[width=0.50\linewidth]{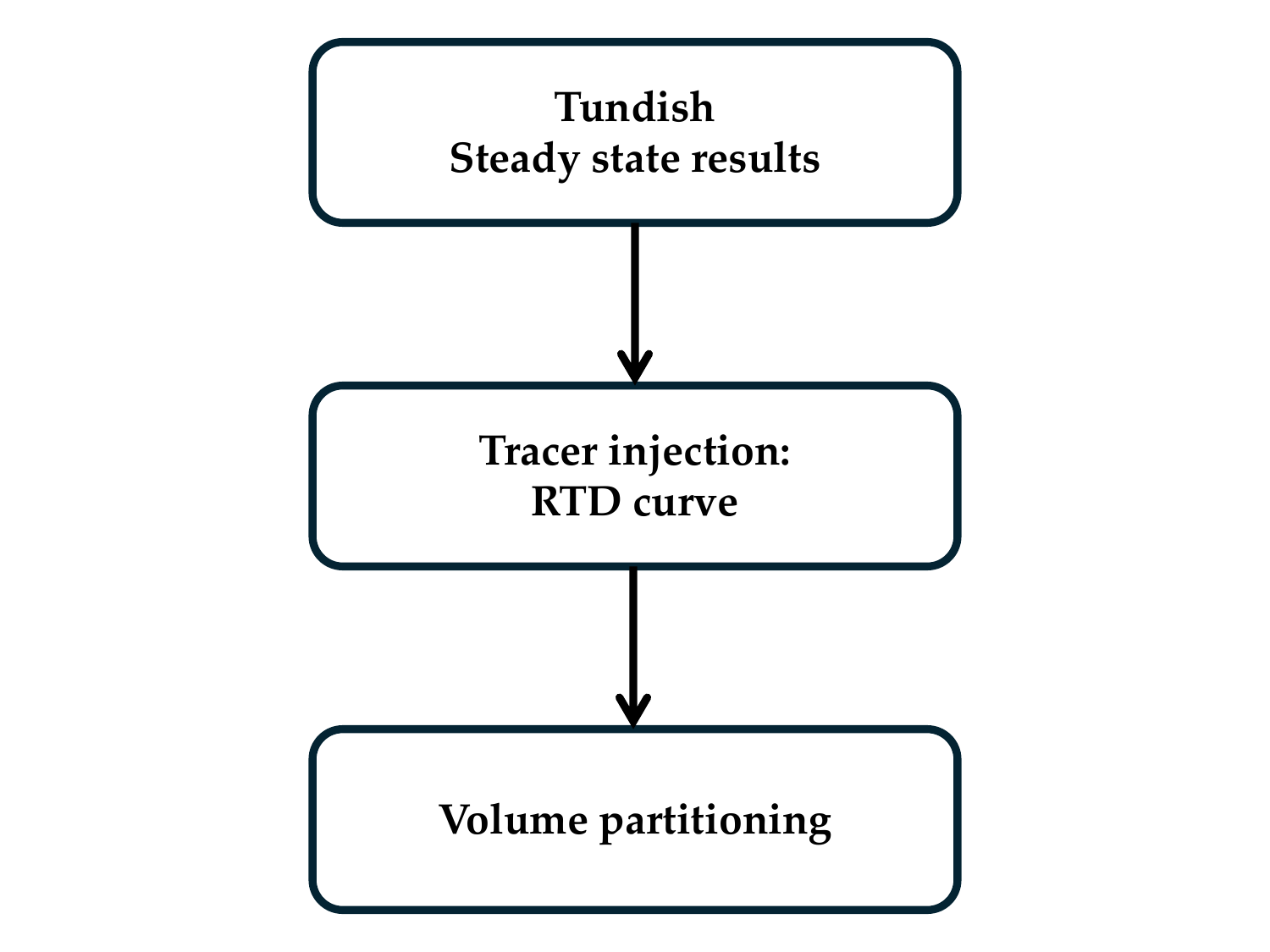} 
    \caption{Schematics of the complete workflow of continuous casting tundish RTD analysis. }
    \label{fig:RTD_workflow}
\end{figure}

As discussed above, the primary objective of this study is to obtain the RTD curve for the continuous casting tundish and, subsequently, to perform RTD analysis for volume partitioning. Figure \ref{fig:RTD_workflow} shows the schematics of the complete workflow of RTD analysis. \\

First, the steady-state flow of the fluid (molten steel or water) is attained for the given initial and boundary conditions. Depending on the fluid considered, the flow may be isothermal (in the case of water) or non-isothermal (for molten steel). A tracer is then injected into the tundish, and its concentration is monitored at the outlet. As reported earlier, the RTD curve is derived from the tracer concentration-time profile recorded at the outlet of the tundish. Next, to perform RTD analysis for volume partitioning, we consider the widely used combined model for the single strand tundish proposed by Sahai and Emi \cite{Sahai1996}. According to the theory, the tundish can be divided into three distinct regions: plug flow, fully mixed flow, and dead flow. The combined plug flow and fully mixed flow region is referred to as the active zone. A tundish can be considered as a combination of different idealized reactors. Sahai and Ahuja \cite{Sahai1986} revised their model, presenting an enhanced combined model, which is as follows:

\begin{equation}
\frac{V_{\text {dead }}}{V}=1-\frac{T_{\text {avg }}}{T_\text{th}},
\label{eqn:deadVolume}
\end{equation}

\begin{equation}
\frac{V_{\text {plug }}}{V}=\frac{\frac{1}{2} (T_{\min } + T_{\max})}{T_\text{th}} = \frac{T_\text{plug}}{T_\text{th}},
\label{eqn:plugVolume}
\end{equation}

\begin{equation}
\frac{V_\text{mix}}{V} = 1 - \frac{V_{\text {dead }}}{V} - \frac{V_{\text {plug }}}{V},
\label{eqn:mixVolume}
\end{equation}

\begin{equation}
    T_\text{th} = \frac{V}{Q}.
    \label{eqn:Tth}
\end{equation}

Where \(T_{\text {avg }}\) is the average residence time, defined as the time required for half of the tracer to exit the tundish; \(T_{\text {th}}\), the theoretical residence time (\ref{eqn:Tth}) is defined as the ratio of the tundish volume \(V\) and the volumetric flow rate \(Q\); \(T_\text{min}\) is the time at which the tracer concentration first reaches \(1 \times 10^{-4} \); and \(T_{\max}\) is the time at which highest concentration is observed. \\

The dead volume fraction (\ref{eqn:deadVolume}) is the region of the tundish in which molten steel remains for an extended duration, typically exceeding twice the theoretical residence time. The plug volume fraction (\ref{eqn:plugVolume}) is the ratio between the average time delay to reach the maximum concentration at the outlet of the tundish and the theoretical residence time, and it's the fraction of active volume where the molten steel flows in a stable manner. The mixing volume fraction (\ref{eqn:mixVolume}) is the fraction of the active tundish volume where steel is homogenized. \\

In this study, we focus on mathematical modelling and water-model experiments of a single-strand tundish to analyze RTD characteristics. The numerical investigation involves simulating three-dimensional, steady-state, turbulent flow under both isothermal (water as working fluid) and non-isothermal (molten steel as working fluid) conditions. Subsequently, the transient evolution of tracer concentration is modelled to obtain the key quantity of interest (QoI), the RTD curve. Further, we perform RTD analysis for volume partitioning, which provides insights into the flow behaviour within the tundish.
    \section{Full order model}\label{sec:FOM}

In this section, we describe the physical and mathematical modelling approach considered to simulate both isothermal and non-isothermal conditions during steady-state tundish operation. Additionally, we outline the modelling strategy employed to capture the transient evolution of a tracer, which is used to determine the RTD curve.

\subsection{Governing equations}
The isothermal fluid flow within the tundish is modelled by numerically solving the continuity and momentum equations under turbulent flow conditions. The governing equations, expressed in Cartesian coordinates, are as follows:

\begin{equation}
\frac{\partial (\rho u_i)}{\partial x_i} = 0, 
\end{equation}
\begin{equation}
    \rho \left(\frac{\partial u_i u_j}{\partial x_j} \right) = \frac{\partial}{\partial x_i}\left[\mu_{\mathrm{eff}}\left(\frac{\partial u_i}{\partial x_j}+\frac{\partial u_j}{\partial x_i}\right)\right]-\frac{\partial P}{\partial x_i} + f , 
\end{equation}
\begin{equation}
\mu_{\mathrm{eff}}=\mu+\mu_{\mathrm{t}}=\mu+\rho c_\mu \frac{k^2}{\varepsilon}.
\end{equation}
Here, the body force is represented by $f$, and the effective viscosity coefficient, \( \mu_{\text{eff}} \), primarily depends on the turbulence parameters \( k \) and \( \varepsilon \), which are determined by solving the following two-equation turbulence model.
\subsection{Turbulence model}
The eddy viscosity-based standard $k-\varepsilon$ two-equation turbulence model \cite{Launder1972} is considered:
\begin{equation}
\rho u_i \frac{\partial k}{\partial x_i} =\frac{\partial}{\partial x_i}\left(\frac{\mu_{\mathrm{eff}}}{\sigma_k} \frac{\partial k}{\partial x_i}\right)+\mu_{\mathrm{t}} \frac{\partial u_i}{\partial x_j}\left(\frac{\partial u_i}{\partial x_j}+\frac{\partial u_j}{\partial x_i}\right)-\rho \varepsilon, 
\end{equation}
\begin{equation}
\rho u_j \frac{\partial \varepsilon}{\partial x_j} = \frac{\partial}{\partial x_j}\left(\frac{\mu_{\mathrm{eff}}}{\sigma_{\varepsilon}} \frac{\partial \varepsilon}{\partial x_j}\right) + C_1 \mu_{\mathrm{t}} \frac{\varepsilon}{K} \frac{\partial u_j}{\partial x_i}\left(\frac{\partial u_i}{\partial x_j}+\frac{\partial u_j}{\partial x_i}\right)-C_2 \frac{\varepsilon}{K} \rho \varepsilon.
\end{equation}
In the equations above, the subscripts \( i \) and \( j \) represent the three spatial coordinate directions (\( x \), \( y \), and \( z \)), respectively. The values of the empirical constants used in the turbulence model are: \( \sigma_k = 1.00 \), \( \sigma_\varepsilon = 1.30 \), \( C_1 = 1.44 \), \( C_2 = 1.92 \), \( C_\mu = 0.09 \), and \( \text{Pr}_t = 0.90 \).

\subsection{Boussinesq approximation} 

To model a non-isothermal fluid flow in the tundish, the following momentum equation is considered:
\begin{equation}
    \rho \left(\frac{\partial u_i u_j}{\partial x_j} \right) = \frac{\partial}{\partial x_i}\left[\mu_{\mathrm{eff}}\left(\frac{\partial u_i}{\partial x_j}+\frac{\partial u_j}{\partial x_i}\right)\right]-\frac{\partial P}{\partial x_i} + \rho_\text{ref} \beta \Delta T g_i, 
    \label{eqn:momentum-Boussinesq} 
\end{equation}
where the last term accounts for the density difference between the existing fluids in the tundish, the last term is derived using the Boussinesq approximation. This approach is commonly employed in natural convection flows to achieve faster convergence. Instead of modelling fluid density as a temperature-dependent variable throughout the domain, the Boussinesq model assumes constant density in all governing equations, except in the buoyancy term of the momentum equation, where density variation with temperature is retained.
\begin{equation}
    (\rho - \rho_\text{ref})g_i = - \rho_\text{ref} \beta (\Delta T)g_i.
\end{equation}

In this context, \( \rho_{\text{ref}} \) denotes the reference fluid density, and \( \beta \) represents the thermal expansion coefficient. The Boussinesq approximation is applicable when temperature variations are small, specifically when the product \( \beta \Delta T \ll 1 \). For water or molten steel, \( \beta \) is typically on the order of \( 10^{-4} \), and even for temperature differences of 15-20\,°C, \( \beta \Delta T \) remains on the order of \( 10^{-3} \), which satisfies the approximation's validity condition. The transient temperature field within the tundish is subsequently evaluated by solving the energy conservation equation.

\begin{equation}
\frac{\partial\left(\rho C_p T\right)}{\partial t}+\frac{\partial\left(\rho C_p T u_j\right)}{\partial x_j}=\frac{\partial}{\partial x_j}\left(k_{e f f} \frac{\partial T}{\partial x_j}\right).
\end{equation}

\noindent where $\rho$ is the density, $C_p$ the specific heat at constant pressure,  
$T$ the temperature, $u_j$ the velocity component in the $x_j$-direction, and $k_{\text{eff}}$ the effective thermal conductivity. The temperature field is coupled with the velocity field via the buoyancy term in (\ref{eqn:momentum-Boussinesq}). In non-isothermal simulations, the tundish walls and the free surface were considered adiabatic, implying that no heat transfer occurred through these boundaries.

\subsection{Species transport equation}

Using the previously computed flow field, the transient evolution of tracer concentration (expressed in terms of mass fraction) is modelled by solving the species transport equation. 

\begin{equation}
\frac{\partial\left(\rho c_i\right)}{\partial t}+\frac{\partial\left(\rho u_j c_i\right)}{\partial x_i}=\frac{\partial}{\partial x_i}\left(\rho D_{\mathrm{eff}} \frac{\partial c_i}{\partial x_i}\right), 
\label{eqn:sepecies_transport}
\end{equation}

\begin{equation}
D_{\mathrm{eff}}=D_m + D_t =D_m + \frac{\mu_{\mathrm{eff}}}{\text{Pr}_t}. 
\end{equation}

Here, $\rho$ is the density, $c_i$ the mass fraction of species $i$, $u_j$ the velocity component, $D_m$ the molecular diffusivity, $D_t$ the turbulent diffusivity, $\mu_{\mathrm{eff}}$ the effective viscosity, and $\text{Pr}_t$ the turbulent Prandtl number.

\section{Numerical method}\label{sec:num_method}

In this study, the OpenFOAM C++ library is utilized for the finite-volume discretization of partial differential equations. The object-oriented structure of C++ allows the code to closely mirror the mathematical formulation, while providing flexibility for further development and modification \cite{Weller1998}. To model isothermal flow, the steady-state solver for incompressible, turbulent flows, \text{simpleFoam}, is considered. For non-isothermal flow, the steady-state solver for buoyant, turbulent flow of incompressible fluids, \text{buoyantBoussinesqSimpleFoam}, is considered. To model the transient evolution of the tracer using the converged steady-state fields, a modified version of \text{scalarTransportFoam} is used, incorporating an additional turbulent diffusivity term in the scalar transport equation. This solver is referred to as \text{scalarTurbulentTransportFoam}. \\

The solution procedure for both isothermal and non-isothermal turbulent flow is based on the SIMPLE (Semi-Implicit Method for Pressure-Linked Equations) algorithm \cite{Caretto}, as implemented in OpenFOAM using the solvers \text{simpleFoam} and \text{buoyantBoussinesqSimpleFoam}, respectively. This algorithm adopts an iterative procedure to solve the governing equations for momentum, pressure, and, in the case of non-isothermal flow, temperature, thereby ensuring the conservation of mass, momentum, and energy. \\

\begin{table}[ht!]
    \centering
    \renewcommand{\arraystretch}{1.2}
    \begin{tabular}{|>{\centering\arraybackslash}p{7cm}|>{\centering\arraybackslash}p{5cm}|}
        \hline
        \textbf{Term} & \textbf{Scheme} \\
        \hline
        Time derivative & steadyState \\
        Convective term (momentum) & linearUpwindV \\
        Convective term (energy) & linearUpwind \\
        Convective term (turbulence: \(k\), \(\epsilon\)) & upwind \\
        Convective term (species) & linearUpwind \\
        Diffusive term (e.g., viscous, thermal) & linear \\
        Gradient term & cellMDLimited linear \\
        \hline
    \end{tabular}
    \caption{Discretization schemes employed to perform FOM simulation of the tundish steady-state operation.}
    \label{table:discretization_scheme}
\end{table}

Table~\ref{table:discretization_scheme} summarizes the discretization schemes employed to solve the governing equations within the finite volume framework. A second-order linear upwind scheme is used for the convective terms in the energy and species transport equations, while the momentum equation employs a velocity direction based linear upwind scheme (linearUpwindV, second-order). For turbulence quantities \( (k, \epsilon)\), a more robust first-order upwind scheme is applied to ensure numerical stability. Diffusive terms, including viscous and thermal diffusion, are discretized using a second-order linear scheme, and the cellMDLimited linear method is employed for gradient computations to maintain boundedness near steep gradients. For transient tracer simulations, a second-order implicit backward time discretization scheme is employed. Appropriate under-relaxation factors are applied to enhance convergence, with values determined empirically.
    \section{Reduced order model}\label{sec:ROM}

This section outlines the projection-based reduced-order modelling framework employed for the RTD analysis. The process is divided into two main stages: offline and online. The offline stage, described first, involves the construction of the reduced-order model from high-fidelity simulation data. This includes generating snapshots, extracting dominant modes using POD, and deriving a reduced system of equations. The subsequent online stage focuses on the efficient evaluation of this model for new parameter instances, which is accelerated using a regression-based approximation of the operators.

\subsection{Offline stage} 

The FOM described in section \ref{sec:FOM} is typically solved for each parameter value in a finite-dimensional training set $\mathcal{K} = \{\mu^k\}_{k=1}^{N_\mu} \subset \mathcal{P}$. Since the problem is both parameter and time dependent, the collection of snapshots for constructing reduced basis spaces must account for both variations. Consequently, a finite set of discrete time instances $\{t^l\}_{l=1}^{N_t} \subset [0, T]$ is also part of the training set. The total number of snapshots is thus $N_s = N_\mu \cdot N_t$. The snapshot matrix \( \mathcal{S} \) is assembled from the snapshots $\{c(\mu^k, t^l)\}_{k=1, l=1}^{N_\mu, N_t}$, which are full-order solutions of the concentration field, evaluated at $N_h$ discrete points in the computational domain:

\begin{equation}
    \mathcal{S} = [ c(\mu^1, t^1), \ldots, c(\mu^{N_\mu}, t^{N_t})] \in \mathbb{R}^{N_h \times N_s}.
\end{equation}

To construct a reduced basis subspace for the projection of the governing equations, there are several established techniques \cite{quarteroni2015reduced, hesthaven2016certified, rozza2022advanced}, such as POD, proper generalized decomposition (PGD), and the reduced basis (RB) method based on a greedy sampling strategy. In the present work, POD is employed to obtain a reduced basis subspace. Particularly, the nested POD method, where POD is first applied to the temporal domain and subsequently in the parametric space \cite{stabile2018finite, rozza2022advanced}.\\

Let \( c(\mu, t) \) be a scalar or vector-valued function, with a set of realizations \( \{c_i\}_{i=1}^{N_s} \). The POD problem seeks, for each dimension of POD space \( N_{\text{POD}} = 1, \ldots, N_s \), a set of orthonormal basis functions \( \{\varphi_j\}_{j=1}^{N_{\text{POD}}} \) that minimize the approximation error:

\begin{equation}
\min_{\{\varphi_j\}_{j=1}^{N_{\text{POD}}}} \sum_{i=1}^{N_s} \left\| c_i - \sum_{j=1}^{N_{\text{POD}}} \langle c_i, \varphi_j \rangle_{L^2(\Omega)} \varphi_j \right\|^2_{L^2(\Omega)}.
\label{eqn:POD_minimization}
\end{equation}
The basis functions must satisfy the orthonormality constraint:
\begin{equation}
\langle \varphi_i, \varphi_j \rangle_{L^2(\Omega)} = \delta_{ij}, \quad \forall i, j = 1, \ldots, N_{\text{POD}}.
\end{equation}
As demonstrated in \cite{quarteroni2015reduced}, the minimization problem presented in (\ref{eqn:POD_minimization}) is mathematically equivalent to solving the associated eigenvalue problem: 
\begin{equation}
\boldsymbol{\mathcal{C}} \boldsymbol{Q} = \boldsymbol{Q} \boldsymbol{\Lambda},
\end{equation}
with the correlation matrix \(\boldsymbol{\mathcal{C}} \in \mathbb{R}^{N_s \times N_s}\) defined as:
\begin{equation}
\boldsymbol{\mathcal{C}}_{ij} = \langle c_i, c_j\rangle_{L^2(\Omega)} \quad \forall \hspace{1mm} i, j=1, \ldots, N_s.
\end{equation}

where \(\boldsymbol{Q}\) is the square matrix containing the eigenvectors as columns, and \(\boldsymbol{\Lambda}\) is the diagonal matrix of corresponding eigenvalues \(\{\lambda_i\}_{i=1}^{N_s}\). The basis functions are subsequently derived as:

\begin{equation}
\varphi_i = \frac{1}{\sqrt{\lambda_i}} \sum_{j=1}^{N_s} c_j (\boldsymbol{Q})_{ij}.
\end{equation}

The POD basis functions obtained using the aforementioned methodology span the following reduced subspace:
\begin{equation}
    V_{rb} = \text{span} \{\varphi_i\}_{i=1}^{N_{rb}} \subset L^2(\Omega),
\end{equation}
where the cardinality \( N_{rb} \ll N_h\) is chosen based on the eigenvalue decay of the associated problem. \\

Once the reduced subspace is constructed, the reduced tracer concentration fields \( c(\mu, t)\) are approximated with:

\begin{equation}
    c(\mu, t) \approx \sum_{i=1}^{N_{rb}} a_i(\mu, t) \varphi_i(x),
    \label{eqn:linearapproximation}
\end{equation}

where \(\{a_i(\mu, t)\}_{i=1}^{N_{rb}}\) are the parameter-time dependent coefficients, and \(\{\varphi_i(x)\}_{i=1}^{N_{rb}}\) are the parameter independent basis functions. \\

The unknown modal coefficients are obtained via Galerkin projection \cite{Lorenzi2016, stabile2018finite} of the governing species transport equation (\ref{eqn:sepecies_transport}) \cite{Khamlich2023} onto a reduced subspace spanned by the POD basis functions. This results in a parameter-dependent reduced-order system of ordinary differential equations (ODEs):
\begin{equation}
\boldsymbol{M}_{\boldsymbol{r}} \dot{\boldsymbol{a}}(t; \mu) + \boldsymbol{C}_{\boldsymbol{r}}(\mu) \boldsymbol{a}(t; \mu) = \boldsymbol{B}_{\boldsymbol{r}}(\mu) \boldsymbol{a}(t; \mu),
\label{eqn:reducedODE}
\end{equation}

where \(\boldsymbol{a}(t; \mu) \in \mathbb{R}^{N_{\text{rb}}}\) denotes the vector of reduced modal coefficients for a given parameter \(\mu \in \mathcal{P}\). These reduced operators are obtained through Galerkin projection of the full-order operators onto the reduced POD subspace:

\begin{equation}
\left\{
\begin{aligned}
\left(\boldsymbol{M}_{\boldsymbol{r}}\right)_{ij} &= \left\langle \varphi_i, \varphi_j \right\rangle_{L^2(\Omega)}, \\
\left(\boldsymbol{B}_{\boldsymbol{r}}(\mu)\right)_{ij} &= \left\langle \varphi_i, D_{\text{eff}}(\mu) \Delta \varphi_j \right\rangle_{L^2(\Omega)}, \\
\left(\boldsymbol{C}_{\boldsymbol{r}}(\mu)\right)_{ij} &= \left\langle \varphi_i, \nabla \cdot (u(\mu) \varphi_j) \right\rangle_{L^2(\Omega)},
\end{aligned}
\right.
\label{eqn:reducedOperators}
\end{equation}

where \(\{\varphi_i\}_{i=1}^{N_{\text{rb}}}\) denote the POD basis functions. Note that the effective diffusivity \(D_{\text{eff}}(\mu)\) and the velocity field \(u(\mu)\) depend on the parameter \(\mu\), thus inducing parametric dependence in the reduced operators. The mass matrix \(\boldsymbol{M_r}\) is parameter-independent as the basis functions are fixed. \\

The POD basis is computed using the method of snapshots \cite{Sirovich1987}. The basis generation and the assembly of the reduced operators (\ref{eqn:reducedOperators}) are performed within the ITHACA-FV framework\footnote{The ITHACA-FV source code is available at \texttt{https://github.com/ITHACA-FV/ITHACA-FV}.} \cite{ITHACA-FV}, an OpenFOAM-based platform for reduced-order modeling of parametrized problems. The resulting system is time-integrated using the implicit backward Euler method, and the full-order solution is reconstructed via the linear modal expansion defined in (\ref{eqn:linearapproximation}).

\subsection{Online stage}

In the online stage of the parametric reduced-order modeling framework, the goal is to rapidly evaluate the reduced system for a new, unseen parameter value \(\mu^\star \in \mathcal{P}\), without solving the full-order problem. To this end, the reduced operators \(\boldsymbol{B}_{\boldsymbol{r}}(\mu^\star)\) and \(\boldsymbol{C}_{\boldsymbol{r}}(\mu^\star)\) must be estimated. The governing reduced-order ODE system is:

\begin{equation}
\boldsymbol{M}_{\boldsymbol{r}} \dot{\boldsymbol{a}}(t; \mu^\star) + \boldsymbol{C}_{\boldsymbol{r}}(\mu^\star) \boldsymbol{a}(t; \mu^\star) = \boldsymbol{B}_{\boldsymbol{r}}(\mu^\star) \boldsymbol{a}(t; \mu^\star).
\label{eqn:onlineODE}
\end{equation}

Since the mass matrix \(\boldsymbol{M}_{\boldsymbol{r}}\) is parameter-independent, it is computed once during the offline stage.
To enable operator prediction at new \(\mu^\star\), we construct regression models to approximate the mappings:

\begin{equation}
\mu \in \mathcal{P} \mapsto \boldsymbol{B}_{\boldsymbol{r}}(\mu), \quad \mu \in \mathcal{P} \mapsto \boldsymbol{C}_{\boldsymbol{r}}(\mu),
\end{equation}
We use regression models trained on the operator values computed at a finite set of training parameters \(\{\mu^k\}_{k=1}^{N_{\text{train}}}\) during the offline stage. Several interpolation methods can be employed for operator approximation \cite{rozza2022advanced}, including inverse distance weighting, Gaussian process regression, or neural networks \cite{Khamlich2023}. This operator approximation is crucial for industrial deployment, where a pre-computed ROM must respond to new parameter queries in real-time without needing access to the original high-fidelity simulation environment.
\\ 

In this work, we use Radial Basis Function (RBF) interpolation to approximate the operators at \(\mu^\star\). The interpolated operators \(\boldsymbol{B}_{\boldsymbol{r}}(\mu^\star)\) and \(\boldsymbol{C}_{\boldsymbol{r}}(\mu^\star)\) are then used to assemble and solve Eq.~(\ref{eqn:onlineODE}) using the backward Euler method. To obtain the reduced solution, the modal coefficients \(\boldsymbol{a}(t; \mu^\star)\) are used to reconstruct the full state via the linear modal expansion:

\begin{equation}
c^{\text{ROM}}(x, t; \mu^\star) = \sum_{i=1}^{N_{\text{rb}}} a_i(t; \mu^\star) \varphi_i(x).
\end{equation}

Although standard RBF interpolation does not guarantee the preservation of structural properties such as symmetry or anti-symmetry of the reduced operators, in our numerical experiments, we observed that the interpolated operators still yield reduced-order solutions that closely match the FOM results across the tested parameter space. This suggests that, in our case, the effect of structure violation is negligible regarding overall accuracy and stability. Nevertheless, this remains a known theoretical limitation, and future extensions could incorporate structure-preserving interpolation strategies to ensure consistency with operator properties. This is further supported by the numerical experiments in section~\ref{sec:results}, where the RBF-interpolated operators' ROM predictions reproduce key dynamics of the full-order system.

    \section{Numerical setup}\label{sec:num_setup}

This section describes the numerical setup considered for simulating the full-order model. The geometry and mesh characteristics of the tundish system are outlined in \ref{subsec:computational_domain}. The boundary conditions applied for both isothermal and non-isothermal simulations are specified in \ref{BCs}.

\subsection{Computational domain}\label{subsec:computational_domain}

\begin{figure}[ht!]
    \centering
    \includegraphics[width=0.75\linewidth]{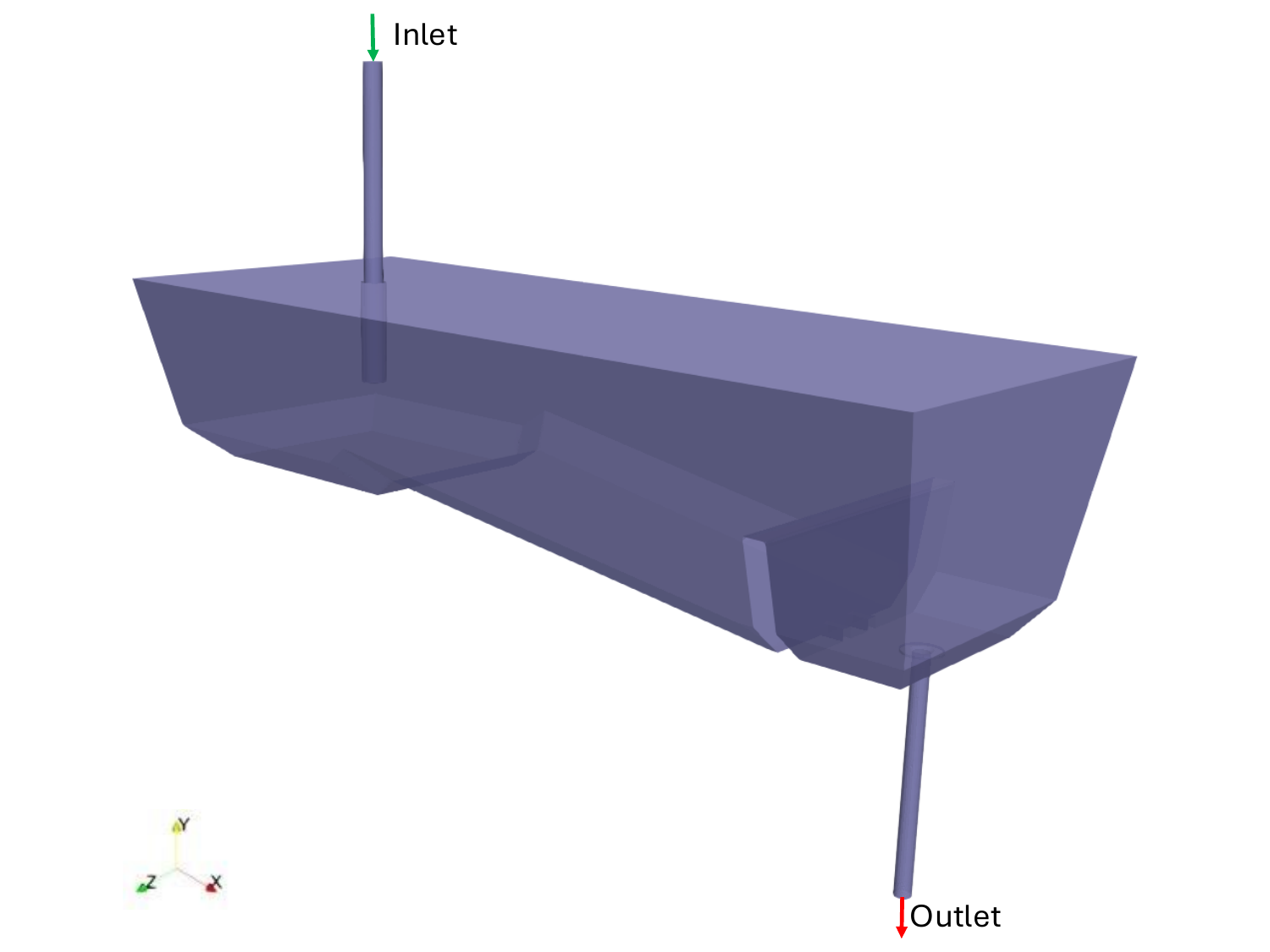}
    \caption{Computational domain of a 1:1 scaled single-strand tundish (isometric view) with specified boundary conditions: inlet for molten steel entry, outlet for steel discharge, and wall boundaries representing refractory-lined surfaces and free surface.}
    \label{fig:geometry}
\end{figure}

A three-dimensional computational domain was constructed based on a \(1{:}1\) scaled industrial tundish, as illustrated in Figure~\ref{fig:geometry}. The inlet, outlet, and remaining surfaces, all treated as walls, are highlighted. \\

\begin{figure}[ht!]
    \centering
    \includegraphics[width=1\linewidth]{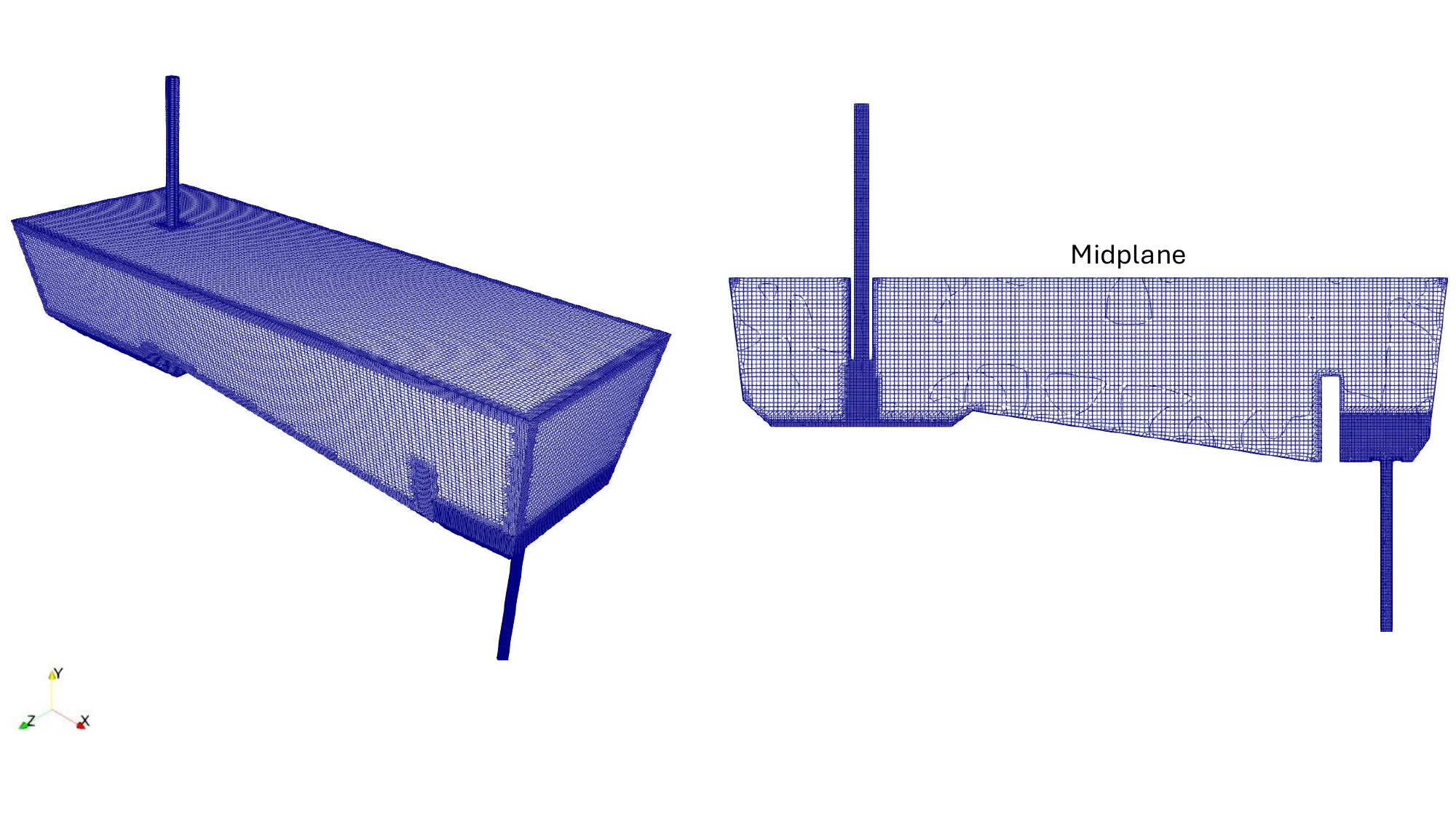}
    \caption{Discretized computational domain of a 1:1 scaled single-strand tundish. \textbf{Left}: Isometric view of the discretized domain; \textbf{Right}: 2D midplane showing mesh refinement at the inlet jet impingement region and outlet, and consists of six prism layers adjacent to the walls.}
    \label{fig:mesh}
\end{figure}

The computational mesh was generated using \text{snappyHexMesh} in OpenFOAM. The final mesh consists of approximately 3.6 million cells, comprising primarily hexahedra (96.8\%), with an additional 6 prism layers, pyramids, tetrahedra, and polyhedra near complex geometrical features, shown in Figure \ref{fig:mesh}.


\subsection{Boundary conditions}\label{BCs}

A constant volumetric flow rate is imposed at the inlet, an outflow boundary condition is applied at the outlet, and no-slip boundary condition is imposed on all walls; the near-wall region is modelled using standard wall functions. For non-isothermal simulation, where the working fluid is molten steel, the heat losses through the lateral and bottom walls are set to \(6~\mathrm{kW}\,\mathrm{m}^{-2}\) and on the free surface . A summary of the input parameters and boundary conditions is provided in Table \ref{table:inputparameters}. \\

The two simulation cases presented in Table~\ref{table:inputparameters} serve complementary purposes. The isothermal water case is included because, at typical casting temperatures (~1600 °C), the kinematic viscosity of molten steel closely matches that of water at room temperature. By preserving dynamic similarity through a constant Froude number, a scaled water model can reproduce the essential flow features of an industrial tundish. This offers a convenient and experimentally accessible benchmark for validating the numerical model. \\

Second, a non-isothermal simulation is performed for an industry-representative molten steel case, where buoyancy effects due to modest temperature gradients are captured using the Boussinesq approximation. These effects play a crucial role in predicting residence time and flow stratification under industrial operating conditions. Together, the two cases serve complementary purposes: the isothermal configuration facilitates validation against experimental measurements, while the non-isothermal case demonstrates the ROM framework's applicability to thermally driven flows in practical settings. \\

\begin{table}[ht!]
\centering
\begin{tabular}{|c|c|c|}
\hline
 & \textbf{Water} & \textbf{Molten steel} \\
\textbf{Parameter} & \textbf{(Isothermal)} & \textbf{(Non-isothermal)} \\
\hline
Density & $1000~\mathrm{kg}/\mathrm{m}^3$ & $6900~\mathrm{kg}/\mathrm{m}^3$ \\
Kinematic viscosity & $1 \times 10^{-6}~\mathrm{m^2}/\mathrm{s}$ & $0.8 \times 10^{-6}~\mathrm{m^2}/\mathrm{s}$ \\
Reference pressure & $101{,}325~\mathrm{Pa}$ & $101{,}325~\mathrm{Pa}$ \\
Heat capacity & -- & $800~\mathrm{J}/\mathrm{kg}\cdot\mathrm{K}$ \\
Thermal conductivity & -- & $35~\mathrm{W}/\mathrm{m}\cdot\mathrm{K}$ \\
Thermal expansion coefficient & -- & $0.000127~\mathrm{K}^{-1}$ \\
Inlet volumetric flow rate  & $0.0102657~\mathrm{m^3}/\mathrm{s}$ & $0.0102657~\mathrm{m^3}/\mathrm{s}$ \\
Inlet velocity & $1.6~\mathrm{m}/\mathrm{s}$ & $1.6~\mathrm{m}/\mathrm{s}$  \\
Inlet temperature & -- & $1823.15~\mathrm{K}$ \\
Wall (flow) & No-slip & No-slip \\
Wall (heat loss) & -- & $6~\mathrm{kW}/\mathrm{m}^2$ \\
\hline
\end{tabular}
\caption{Input parameters and boundary conditions used for the isothermal and non-isothermal simulations.}
\label{table:inputparameters}
\end{table}

\begin{figure}[ht!]
    \centering
    \includegraphics[width=0.65\linewidth]{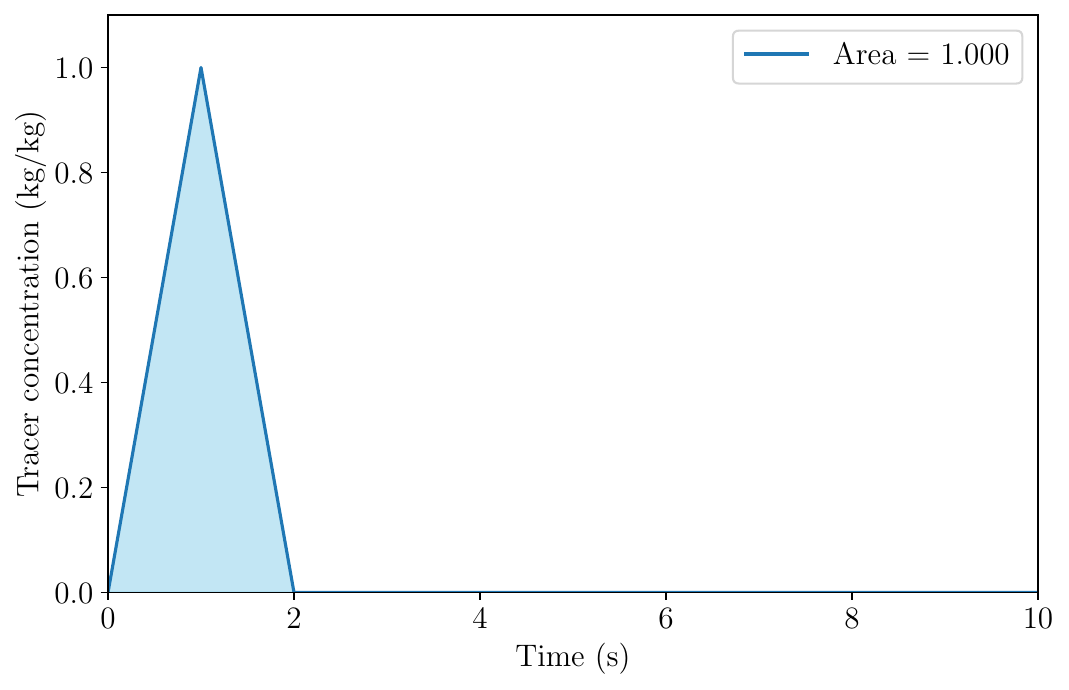}
    \caption{Pulse input of tracer concentration at the inlet boundary used for the transient simulation.}
    \label{fig:Tracer_inletBC}
\end{figure}

As shown in Figure~\ref{fig:Tracer_inletBC}, the tracer concentration \( c \), expressed as a mass fraction (kg/kg), is set to unity at the inlet for \( 0 \leq t \leq 2~\mathrm{s} \), and zero thereafter. The outlet tracer concentration is monitored over \( 0 \leq t \leq 3000~\mathrm{s} \) to construct the RTD curves.

    \section{Results and discussion}\label{sec:results}
\begin{figure}
    \centering
    \includegraphics[width=0.75\linewidth]{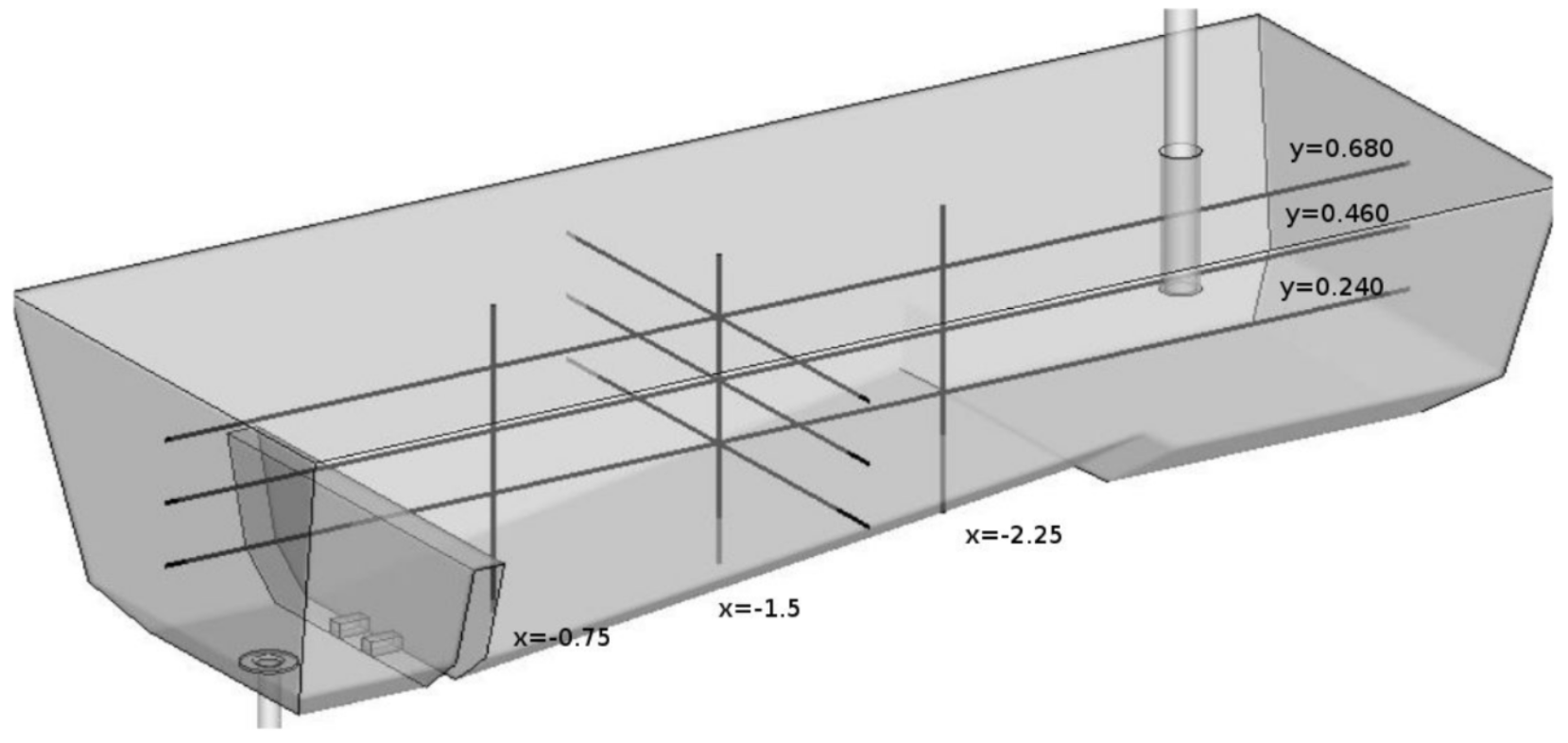}
    \caption{Locations of horizontal and vertical lines used for extracting streamwise velocity profiles within the tundish.}
    \label{fig:tundish_Vlines}
\end{figure}

\begin{figure}[ht!]
    \centering
    \includegraphics[width=1\linewidth]{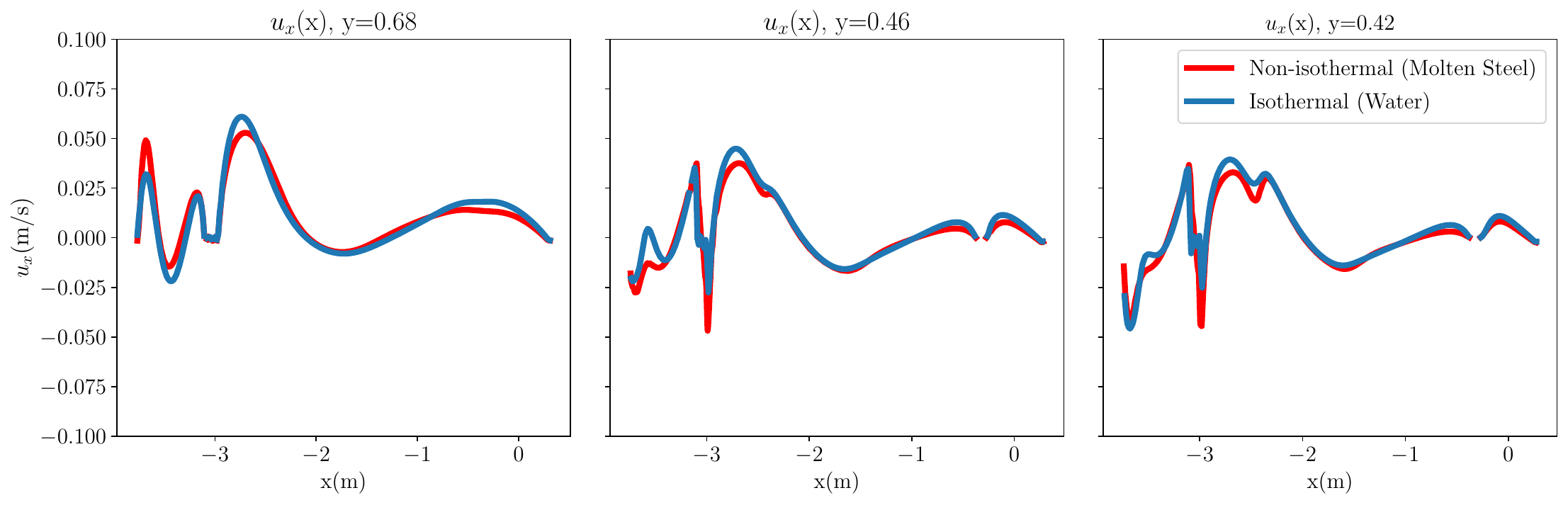}
    \caption{Streamwise velocity distributions along horizontal lines at different y-positions (shown in Figure \ref{fig:tundish_Vlines}), under isothermal and non-isothermal conditions.}
    \label{fig:velocity-horizontal-line}
\end{figure}

\begin{figure}[ht!]
    \centering
    \includegraphics[width=1\linewidth]{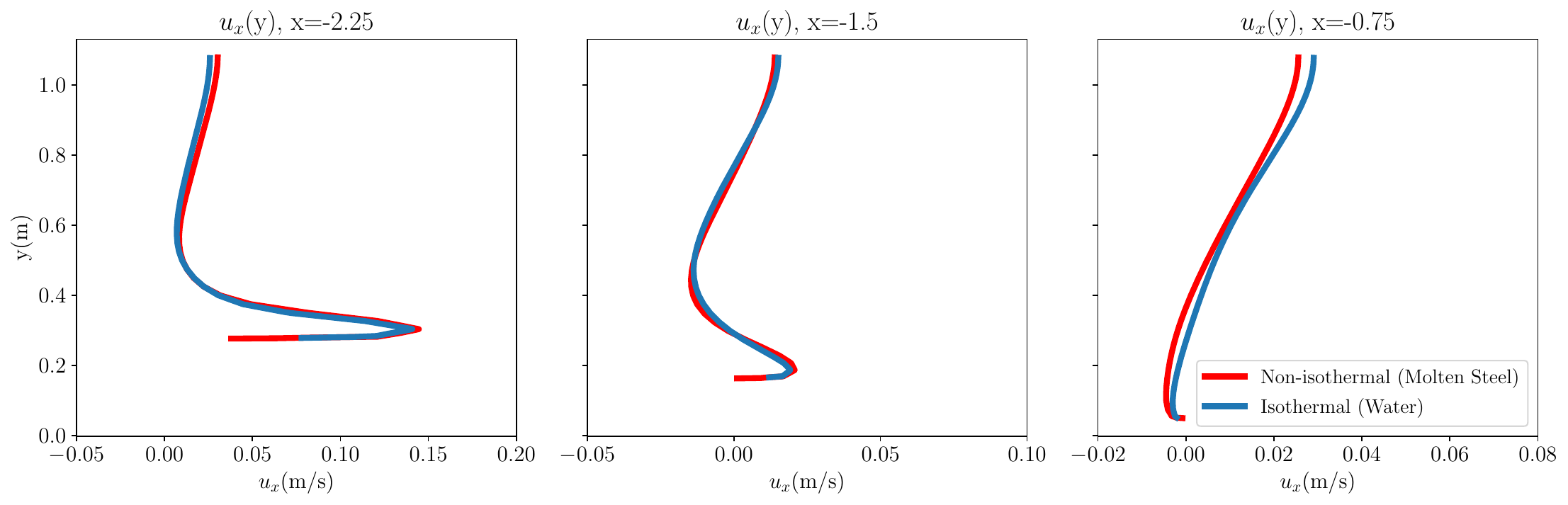}
    \caption{Streamwise velocity distributions along vertical line at different x locations (shown in Figure \ref{fig:tundish_Vlines}), under isothermal and non-isothermal conditions.}
    \label{fig:velocity-vertical-line}
\end{figure}

In this work, we perform three-dimensional steady-state turbulent flow simulations under both isothermal (where water is the working fluid) and non-isothermal (where molten steel is the working fluid) conditions. The streamwise velocity component \(u_x\) is extracted along horizontal and vertical lines (shown in Figure \ref{fig:tundish_Vlines}) within the tundish to assess the lateral and vertical variations in the flow field. Figure \ref{fig:velocity-horizontal-line} presents a comparison of the streamwise velocity distributions under isothermal and non-isothermal conditions along horizontal lines located at various y-positions. Figure \ref{fig:velocity-vertical-line} shows the vertical variation of the streamwise velocity along different x-positions. \\


It is observed that the effect of buoyancy under non-isothermal conditions on the velocity field is negligible, as the velocity distributions under isothermal and non-isothermal conditions match closely. These velocity distributions, together with the tracer studies, provide the basis for the subsequent RTD analysis and volume partitioning, enabling a comprehensive assessment of the flow behaviour and mixing efficiency within the tundish. \\

\begin{figure}[ht!]
    \centering
    \includegraphics[width=0.7\linewidth]{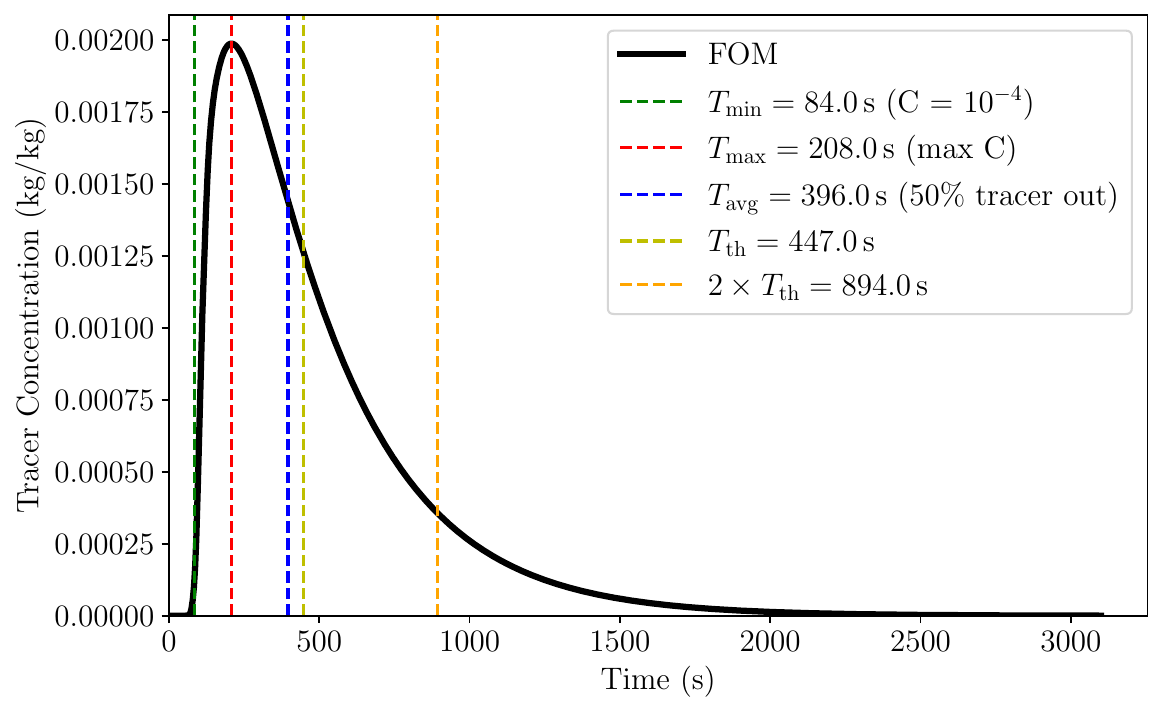}
    \caption{RTD curve obtained from FOM CFD simulations, highlighting key flow characteristics.}
    \label{fig:RTD-analysis}
\end{figure}

\begin{figure}[ht!]
    \centering
    \includegraphics[width=0.6\linewidth]{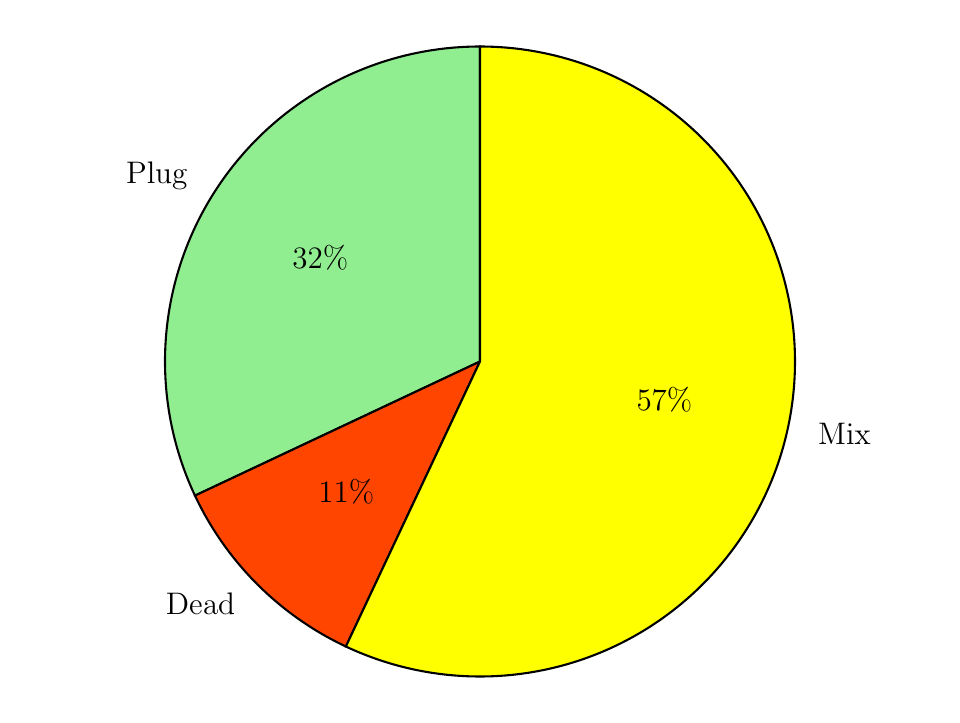}
    \caption{Volume partitioning derived from RTD analysis based on FOM simulation results.}
    \label{fig:volume_partition}
\end{figure}

The transient simulation of tracer injection is employed to characterize fluid flow behaviour within the tundish. The resulting tracer concentration-time profile, referred to as the RTD curve, forms the basis for RTD analysis and volume partitioning. Figure \ref{fig:RTD-analysis} presents the RTD curve obtained from the FOM CFD simulations, with the key flow characteristics such as \(T_{\min}, T_{\max}, T_{\text{avg}}, T_{\text{th}}\) of the RTD curve being highlighted. The resulting plug volume \(V_{\text{plug}}\), mixing volume \(V_{\text{mix}}\), and dead volume \(V_{\text{dead}}\) are presented in Figure \ref{fig:volume_partition}. \\


Table \ref{Table:RTD_parameters} presents the inlet velocities and flow rates used in the parametric study of the RTD curves for the single-strand tundish. These values are considered to perform a sensitivity analysis on the RTD characteristics of the system. \\

\begin{table}[ht!]
    \centering
    \renewcommand{\arraystretch}{1}
    \begin{tabular}{|>{\centering\arraybackslash}p{3cm}|
                    >{\centering\arraybackslash}p{4cm}|
                    >{\centering\arraybackslash}p{4cm}|}
        \hline
        \textbf{Parameter No.} & \textbf{Flow rate (\si{m^3/hr})} & \textbf{Velocity (\si{m/s})} \\
        \hline
        1 & 36.9365 & 1.6301 \\
        2 & 34.52985 & 1.5239 \\
        3 & 32.1232 & 1.4177 \\
        4 & 29.71655 & 1.3115 \\
        5 & 27.3099 & 1.2053 \\
        6 & 24.90325 & 1.0991 \\
        7 & 22.4966 & 0.9929 \\
        8 & 20.08995 & 0.8866 \\
        9 & 17.6833 & 0.7804 \\
        10 & 15.27665 & 0.6742 \\
        11 & 12.87 & 0.5680 \\
        \hline
    \end{tabular}
    \caption{Inlet velocities and flow rates used in the parametric study of the RTD curves for the single-strand tundish. These parameters were employed to perform a sensitivity analysis on the RTD characteristics.}
    \label{Table:RTD_parameters}
\end{table}

\begin{figure}[ht!]
    \centering
    \includegraphics[width=0.8\linewidth]{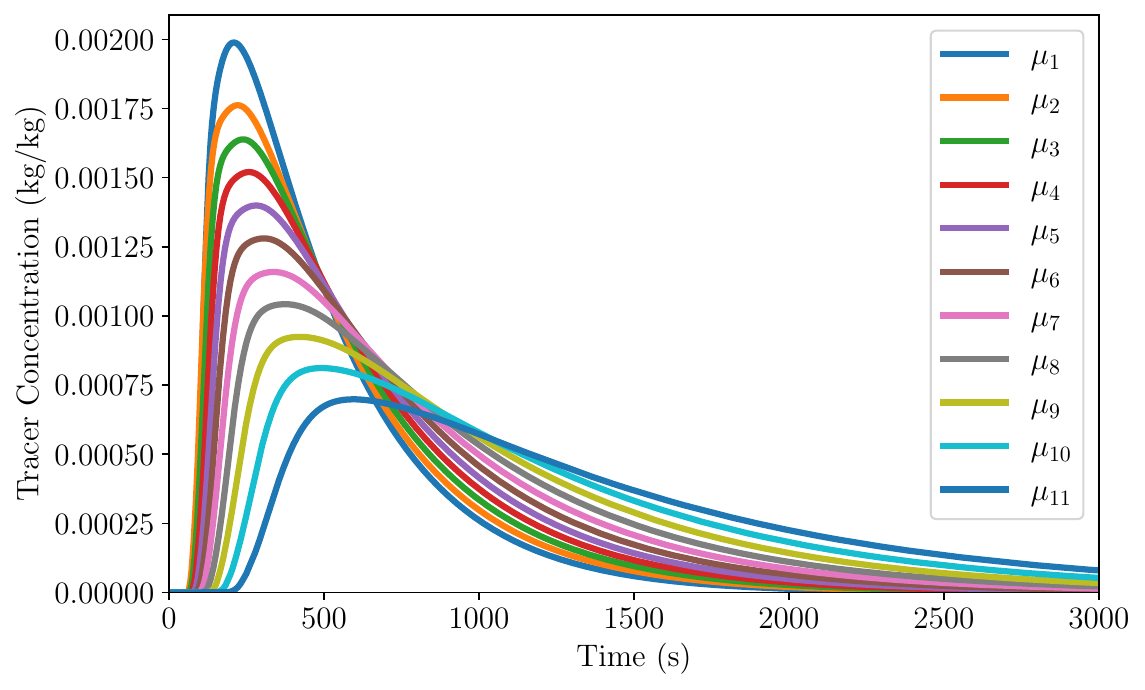}
    \caption{RTD curves obtained from full-order simulation, illustrating the variation in tracer concentration over time for each parameter, as listed in Table \ref{Table:RTD_parameters}.}
    \label{fig:RTD_FOM_results}
\end{figure}

For the parametric study of the RTD curves, we first obtain the convective velocity field by performing isothermal steady-state simulations for each parameter listed in Table \ref{Table:RTD_parameters}. Next, transient scalar transport simulations are carried out to derive the corresponding RTD curves, which are presented in Figure \ref{fig:RTD_FOM_results}. As the inlet velocity decreases, the RTD curves broaden, indicating an increase in fluid residence time within the tundish. The peak concentration of the RTD curve also decreases, and the distribution becomes more spread out. This occurs because lower velocities result in slower fluid movement, leading to less efficient mixing and a wider range of residence times to exit the system. Consequently, the RTD curves at lower velocities are less sharp and more spread out, highlighting the impact of velocity on residence time distribution and system efficiency. These changes highlight the influence of velocity on RTD characteristics and system performance. \\

In this study, the projection-based ROM is constructed following the procedure described in section \ref{sec:ROM}. To construct the parametric projection-based ROM for RTD analysis, we consider parameter values corresponding to distinct simulation cases with varying inlet velocity conditions. Specifically, we select eight parameter instances indexed as \( \mu_i, \text{where}, i= \{ 1, 2,3,4,7,8,9,10\} \), form the training set (from Table \ref{Table:RTD_parameters}). POD is performed on the training set, the first 21 modes capturing 99.99$\%$ of the cumulative energy, are retained to construct the reduced basis subspace. Subsequently, the governing equations for tracer transport are projected onto the reduced basis subspace using Galerkin projection. Following the projection, a reduced-order system of ODEs is obtained, governing the temporal evolution of the modal coefficients. These ODEs are integrated in time using an implicit backward Euler method. The reduced tracer concentration fields are then reconstructed by a linear combination of the time-dependent coefficients and the spatial POD modes. To obtain the parameter-dependent reduced operator in the online phase, we build the parameter-to-reduced operator mapping using RBF interpolation and employ it in the online phase for fast evaluation of reduced operators \( \boldsymbol{B}_{\boldsymbol{r}}(\mu) \) \text{and} \(\boldsymbol{C}_{\boldsymbol{r}}(\mu)\). The remaining parameter instances, \(\mu_i, \text{where}, i= \{5,6,11\}\), are for testing. These test parameters are not used during the training phase and serve to evaluate the predictive capabilities and generalization of the ROM, especially for interpolation and mild extrapolation regimes. \\

\begin{figure}[ht!]
    \centering
    \includegraphics[width=0.72\linewidth]{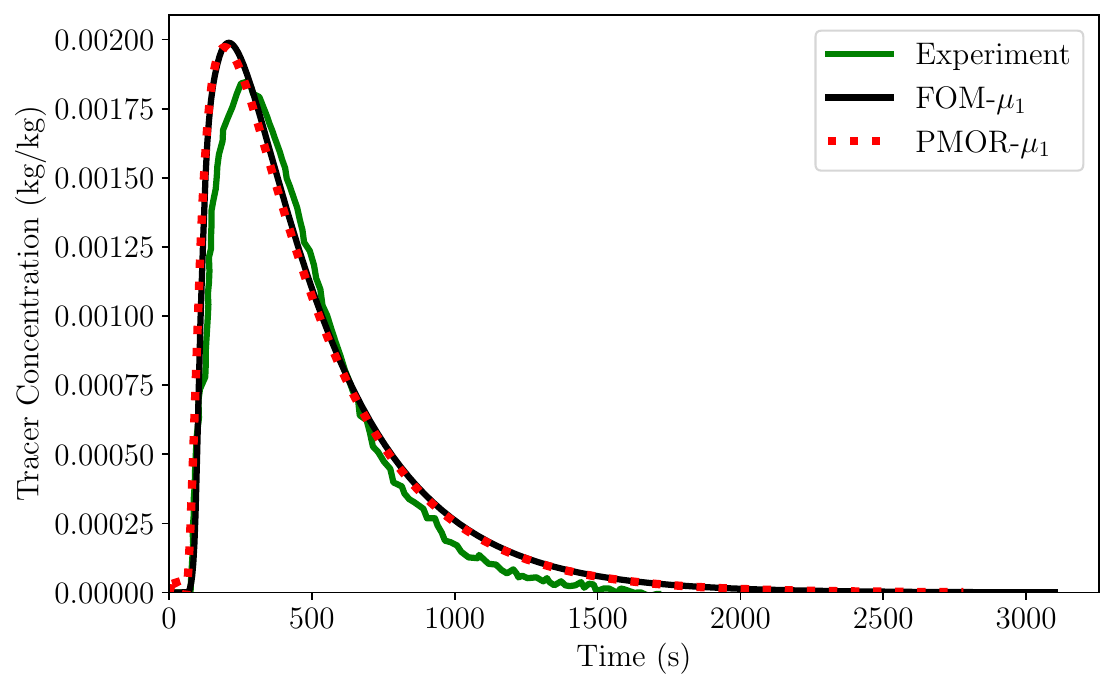}
    \caption{Comparison of RTD curves from the experiment, FOM CFD simulation, and projection-based ROM, illustrating the good agreement between the simulations and experimental data.}
    \label{fig:EXP_FOM_ROM_prediction}
\end{figure}

Figure~\ref{fig:EXP_FOM_ROM_prediction} compares the RTD curves for parameter $\mu_1$ obtained from the experiment, the FOM CFD simulation, and the projection-based ROM. Both the FOM and ROM predictions exhibit good agreement with the experimental data, with only minor discrepancies observed. \\ 

In the interpolatory regime, \(\mu_i, \text{where}, i= \{5,6\}\), the parametric projection-based ROM shows excellent agreement with the full-order simulation, accurately predicting the RTD curves for test parameters within the range used for training. The model captures the dynamics of the system effectively, providing reliable results for these parameters. \\

\begin{figure}[ht!]
    \centering
    \includegraphics[width=0.7\linewidth]{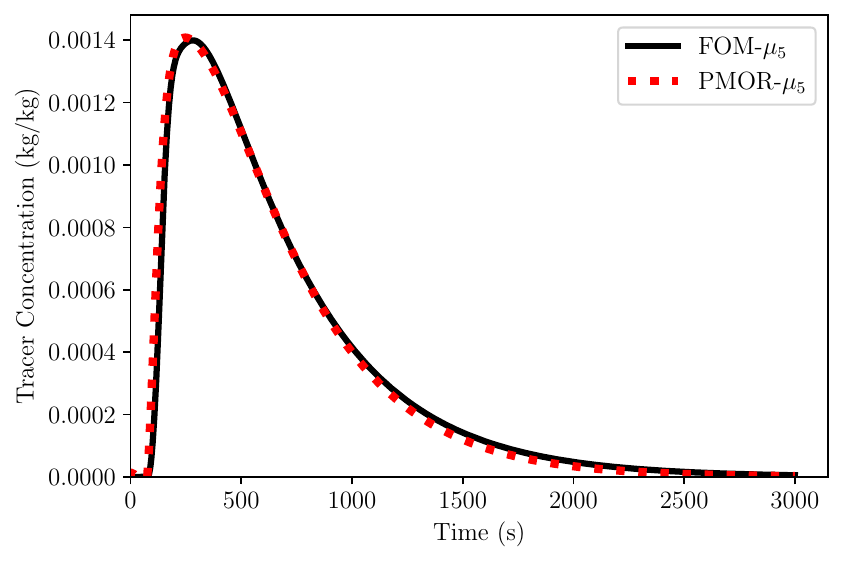}
    \caption{RTD curves for test parameter $\mu_5$: comparison between the parameter-time dependent projection-based ROM and full-order simulation results, showing excellent agreement within the interpolatory regime.}
    \label{fig:PMOR_prediction_1}
\end{figure}

\begin{figure}[ht!]
    \centering
    \includegraphics[width=0.7\linewidth]{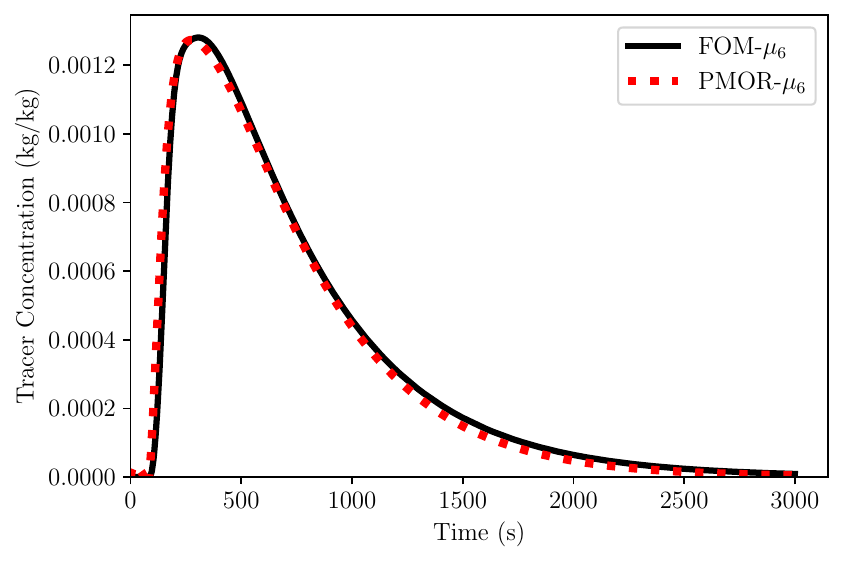}
    \caption{RTD curves for test parameter $\mu_6$: comparison between the parameter-time dependent projection-based ROM and full-order simulation results, showing excellent agreement within the interpolatory regime.}
    \label{fig:PMOR_prediction_2}
\end{figure}

In the extrapolatory regime, where the test parameter $\mu_{11}$ is outside the training range, the ROM still captures the overall trend of the RTD curve but shows a slight discrepancy in maximum concentration prediction. This demonstrates the ROM's ability to extrapolate beyond the trained parameters, though with some limitations in precision. \\

\begin{figure}[ht!]
    \centering
    \includegraphics[width=0.7\linewidth]{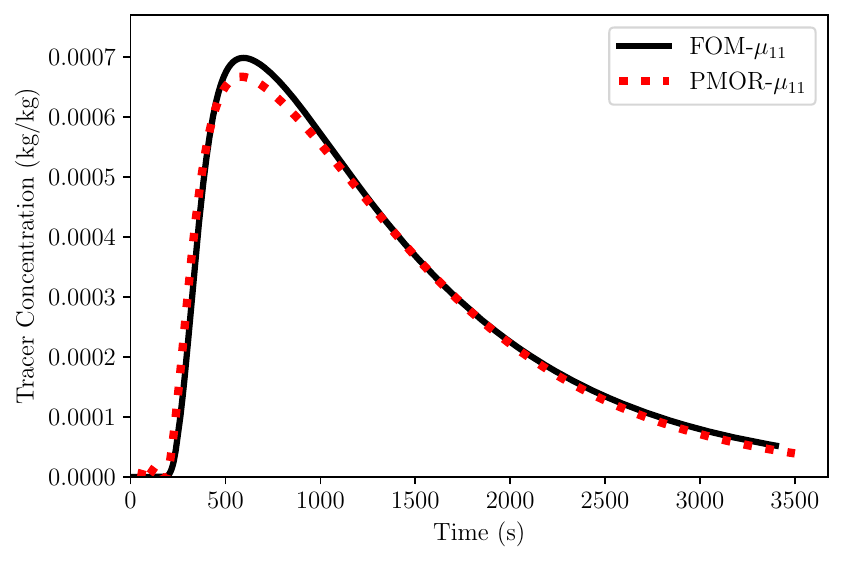}
    \caption{RTD curves for test parameter $\mu_{11}$ in the extrapolatory regime: comparison between the parameter-time dependent projection-based ROM and full-order simulation results, ROM predictions accurately capturing the overall trend of the RTD curve, with a slight discrepancy in the maximum concentration prediction.}
    \label{fig:PMOR_prediction_3}
\end{figure}

The computational resources and wall-clock time required for the FOM simulations are summarised in Table~\ref{tab:steady-state-FOMcomp-resources}. The computational efficiency achieved using the projection-based ROM for evaluating the QoIs is presented in Table~\ref{tab:steady-state-IntrusiveROMcomp-resources}. While individual FOM simulations range approximately from 8.19 and 23.17 hours of wall-clock time on 64 to 40 cores, the ROM’s online evaluation completes in only 0.04 seconds on a single core, resulting in a speed-up of approximately $1 \times {10 ^ 6}$ faster in wall-clock time compared to the average FOM simulation. Although the ROM offline construction demands around 40 hours on 2 cores, comparable to a single FOM simulation, this cost is a one-time investment, enabling rapid repeated evaluations. This substantial acceleration highlights the ROM’s suitability for real-time prediction and optimization tasks.

\begin{table}[ht!]
    \centering
    \renewcommand{\arraystretch}{1.2} 
    \begin{tabular}{|>{\centering\arraybackslash}p{1.7cm}|>{\centering\arraybackslash}p{4.cm}|>{\centering\arraybackslash}p{4cm}|>{\centering\arraybackslash}p{2.8cm}|}        
    \hline
    \textbf{Expt No.} & \textbf{Wall-clock time} & \textbf{CPU time (core-hours)} & \textbf{Cores Allocated} \\     
    \hline
    1 & 2.81 + 5.38  hours &  177.05 + 360 hours    &   64 \\ 
    2 &  4.43 + 9 hours  &  177.2 + 360 hours   &   40 \\ 
    3 &  7.26 + 9 hours &  283.08 + 360 hours   &   40 \\ 
    4 &  6.67 + 9  hours &  260.92 + 360 hours   &   40 \\  
    5 &  5.16 + 9  hours &  201.55 + 360 hours   &   40 \\  
    6 &  4.24 + 9  hours &  165 + 360 hours     &   40 \\  
    7 &  14.17 + 9  hours &  555.24 + 360 hours   &   40 \\  
    8 &  3.75 + 9  hours &  145.53 + 360 hours   &   40 \\  
    9 &  6.10 + 9  hours &  237.70 + 360 hours   &   40 \\  
    10 &  6.54 + 9  hours &  256.27 + 360 hours   &   40 \\  
    11 &  10.48 + 9  hours &  410.90 + 360 hours   &   40 \\  
    \hline
    \end{tabular}
    \caption{Summary of computational resources used for the FOM simulations to obtain the tundish steady-state operation QoIs on the Linux cluster. The wall-clock time refers to the elapsed real time, while the CPU time denotes the cumulative usage across all allocated cores (core-hours). These results correspond to the steady Reynolds averaged Navier-Stokes + transient tracer simulations discussed in Table \ref{Table:RTD_parameters}.}
    \label{tab:steady-state-FOMcomp-resources}
\end{table}

\begin{table}[ht!]
    \centering
    \renewcommand{\arraystretch}{1.2} 
    \begin{tabular}{|>{\centering\arraybackslash}p{1.8cm}|
                    >{\centering\arraybackslash}p{2.2cm}|
                    >{\centering\arraybackslash}p{2.5cm}|
                    >{\centering\arraybackslash}p{2.8cm}|
                    >{\centering\arraybackslash}p{2.cm}|}        
    \hline
    \textbf{Intrusive ROM} & \textbf{Offline Wall-clock Time} & \textbf{Offline CPU Time (core-hours)} & \textbf{Online Time (Wall-clock)} & \textbf{Cores Used (Offline / Online)} \\
    \hline
    PMOR &  40 hours &  80 core-hours & 0.04 seconds & 2 / 1 \\
    \hline
    \end{tabular}
    \caption{Computational resources used for projection-based ROM construction and prediction to obtain the tundish steady-state operation QoIs. The offline phase includes computation of POD modes (whole domain is considered), assembling reduced operators, and solving the reduced-order ODE to obtain the modal coefficients. The online phase refers to the prediction of the desired QoI at the tundish outlet. Here, online evaluation is performed only at the outlet probe location, where the QoI is recorded. Evaluating over the entire domain takes approximately 46 seconds, primarily due to the cost of the dot product operation.}
    \label{tab:steady-state-IntrusiveROMcomp-resources}
\end{table}

    \section{Conclusions and perspectives}\label{sec:conclusion}

Over the past two decades, mathematical modelling has become an essential tool for investigating flow phenomena within steelmaking tundishes. Due to the challenging nature of experimental studies in such high-temperature environments, physical investigations remain limited in scope. Advances in computational techniques have enabled significant progress, with numerical simulations now playing a pivotal role in optimising tundish performance. \\

In this study, full-order simulations were performed under both isothermal and non-isothermal conditions. The results indicate that buoyancy effects under non-isothermal conditions have minimal influence on the velocity field, as the velocity distributions in both cases are nearly identical. These velocity fields, combined with tracer transport simulations, serve as the foundation for the subsequent analysis of RTD and volume partitioning. Together, these results provide a comprehensive assessment of the flow behaviour and mixing efficiency within the tundish. \\

Further, a comparison of RTD curves obtained from experiment, full-order simulation, and the projection-based ROM shows that both the FOM and ROM predictions closely match the experimental measurements, with only minor discrepancies. These results validate the accuracy of the reduced-order model and highlight its potential as an efficient surrogate for full-scale simulations in analysing flow behaviour within the tundish. For the parametric RTD analysis, the projection-based ROM demonstrates excellent accuracy in the interpolatory regime, with predictions closely matching full-order simulations. However, in the extrapolatory regime, slight discrepancies are observed, but the ROM captures the overall trend of the RTD curve. \\

Future work will focus on extending the intrusive ROM frameworks \citep{stabile2018finite, stabile2019reduced, SokratiaGeorgaka2020} to both isothermal and non-isothermal tundish steady-state operations. Furthermore, the framework will be adapted to address more complex industrial scenarios, including multi-outlet tundishes and the integration of advanced structure-preserving interpolation methods for the reduced operators. \\

Overall, the integration of ROM techniques significantly reduces computational costs while retaining essential flow characteristics, demonstrating their potential for real-time analysis, design optimization, and digital twin application in metallurgical processes such as continuous casting.
    \section*{\large CRediT authorship contribution statement}
\textbf{Harshith Gowrachari}: Writing - original draft, Conceptualization, Data curation, Formal Analysis, Investigation, Methodology, Software, Validation, Visualization. \textbf{Mattia Giuseppe Barra}: Writing – original draft and review $\&$ editing, Methodology, Investigation. 
\textbf{Moaad Khamlich}: Writing – review $\&$ editing, Methodology, Software. 
\textbf{Giovanni Stabile}: Writing – review $\&$ editing, Methodology, Software, Supervision. 
\textbf{Gianluca Bazzaro}: Project administration, Supervision.   
\textbf{Gianluigi Rozza}: Funding acquisition, Project administration, Supervision.  

\section*{\large Declaration of Generative AI and AI-assisted technologies in the writing process}
These technologies were used to improve readability and correct spelling during the preparation of this manuscript. After using them, the authors reviewed and edited the content as needed and take full responsibility for the content of the publication. 

\section*{\large Declaration of competing interest}
The authors declare that they have no known competing financial interests or personal relationships that could have appeared to influence the work reported in this paper

\section*{\large Data statement}
Access to the data will be unavailable, as the research data includes sensitive and confidential information. 

\section*{Acknowledgements}
We acknowledge the PhD grant supported by industrial partner Danieli \& C. S.p.A. and Programma Operativo Nazionale Ricerca e Innovazione 2014-2020, P.I. Gianluigi Rozza. HG gratefully acknowledges Mattia Giuseppe Barra, Gianluca Bazzaro and Gabriele Guastaferro for their valuable discussions and for facilitating the hosting arrangements during multiple visits to Danieli Research Center (DRC).  GS acknowledges the financial support under the National Recovery and Resilience Plan (NRRP), Mission 4, Component 2, Investment 1.1, Call for tender No. 1409 published on 14.9.2022 by the Italian Ministry of University and Research (MUR), funded by the European Union – NextGenerationEU– Project Title ROMEU – CUP P2022FEZS3 - Grant Assignment Decree No. 1379 adopted on 01/09/2023 by the Italian Ministry of Ministry of University and Research (MUR) and acknowledges the financial support by the European Union (ERC, DANTE, GA-101115741). Views and opinions expressed are however those of the author(s) only and do not necessarily reflect those of the European Union or the European Research Council Executive Agency. Neither the European Union nor the granting authority can be held responsible for them. 
\bibliographystyle{abbrvnat}
\bibliography{bib/biblio}

\end{document}